\documentstyle[12pt]{article}

\makeatletter \@addtoreset{equation}{section} \makeatother

\def\ads{AdS_{5}}
\def\be{\begin{equation}}
\def\ee{\end{equation}}
\def\ba{\begin{array}}
\def\ea{\end{array}}
\def\d{\partial}
\def\dps{\displaystyle}

\def\ba{\begin{array}}
\def\ea{\end{array}}
\def\d{\partial}
\def\dps{\displaystyle}

\def\bpsi{\bar{\psi}}
\newcommand{\hsa}{$cu(1,0|8)\,\,$}
\newcommand{\half}{\frac{1}{2}}

\begin{document}

\begin{flushright}
FIAN/TD/11/02\\
June 2002\\
\end{flushright}

\vspace{0.5cm}

\begin{center}
{\Large ${\cal N}=1$ Supersymmetric Theory of \\
Higher Spin Gauge Fields in $AdS_5$\\
at the Cubic Level}

\vglue 0.6  true cm \vskip1cm

K.B.~Alkalaev$^{\,\dag}$ and M.A.~Vasiliev$^{\,\ddag}$
\vglue 0.3  true cm

\medskip
{\it I.E.Tamm Department of Theoretical Physics, Lebedev Physical Institute,\\
Leninsky prospect 53, 119991, Moscow, Russia}
\\
\vspace{5mm} {\it $^\dag$ E-mail: {\tt alkalaev@lpi.ru}}  \\ {\it
$^\ddag$ E-mail: {\tt vasiliev@lpi.ru}}

\medskip
\vskip1cm
\end{center}

\begin{abstract}
We formulate gauge invariant  interactions of totally symmetric
tensor and tensor-spinor higher spin gauge fields in $\ads$ that
properly account for higher-spin-gravitational interactions at the
action level in the first nontrivial order.

\end{abstract}


\section{Introduction}
Study of the higher spin theory in $AdS$ background is of interest
due to its potential relationship (see e.g.
\cite{V1,V_obz,V_obz2,Metsaev:1999} and reference therein) with a
symmetric phase of a theory of fundamental interactions presently
identified with M theory. An additional motivation for the study
of higher spin gauge theories came recently
\cite{HSads/cft,SS1,Vc,Wit2,AAT,Sezgin:2002} from somewhat
different arguments based on $AdS/CFT$ correspondence
\cite{JM,GKP,Wit}, pointing at the same direction. {}From this
perspective the case of $\ads$ is of particular importance,
because higher spin gauge theories in $\ads$ are dual to $4d$
superconformal theories. The case of $\cal{N}$=$4$ supersymmetry
is most interesting as the corresponding $4d$ superconformal model
is  $\cal{N}$=$4$ $SYM$.

In \cite{VD5} it has been shown  that totally symmetric bosonic
higher spin gauge fields propagating on $AdS_5$ admit consistent
higher-spin-gravitational interactions at least in the cubic
order. The corresponding action was constructed in the first
nontrivial order. The system exhibits higher spin symmetries
associated with certain  higher spin algebra originally introduced
in \cite{FLA} and called \hsa in \cite{Vc} and requires $AdS$
geometry rather than the flat one thus extending $4d$ results of
\cite{FV1} to $d=5$. One difference compared to the $4d$ case is
that the $5d$ higher spin algebra  \hsa contains non-trivial
center freely generated by the central element $N$ \cite{FLA}. As
a result, \hsa gives rise to the infinite sets of fields of any
spin. The factorization of the algebra \hsa with respect to the
maximal ideal generated by $N$, that gives rise to the reduced
higher spin algebra $hu_0(1,0|8)$ in which every integer spin
appears in one copy, was shown to admit consistent interactions as
well \cite{VD5}.

In this paper we continue the study of higher spin interactions of
totally symmetric massless fields in $AdS_5$, extending the
analysis of \cite{VD5} to the model with fermions that exhibits
the higher spin symmetries associated with the simplest $\ads$
higher spin superalgebra $cu(1,1|8)$. The totally symmetric higher
spin gauge fields originating  from $cu(1,1|8)$ are arranged into
an infinite sequence of supermultiplets $\{s\}^{(k)}, 0\leq k
<\infty$, with a spin content $(s,s-\frac{1}{2},s-1)^{(k)}$
determined by an integer highest spin $s=2,3,...,\infty \;$.
Strictly speaking, the theory we consider is not fully
supersymmetric because we truncate away all lower spin fields with
$s\leq 1$ (in particular, the spin 1 field from the spin 2
supermultiplet). This truncation is done to simplify analysis
because lower spin fields require special formulation while our
goal is to check consistency of the higher-spin-gravitational
interactions. By analogy with the $4d$ analysis (see second
reference in  \cite{FV1}) it is not expected to be a hard problem
to extend our analysis to the case with lower spin fields
included. Note that a truncation of  lower spin fields is only
possible  at the cubic level\footnote{At the cubic level such an
incomplete system remains formally consistent because one can
switch out interactions among any three elementary (i.e.,
irreducible at the free field level) fields without spoiling the
consistency  at this order. This is a simple consequence of the
Noether current interpretation of the cubic interactions: setting
to zero some of the fields is always consistent with the
conservation of currents.}
 and these fields (in particular, scalar fields) have necessarily to be
introduced in the analysis of higher-order corrections.
Correspondingly, we will refer to the theory under consideration
as to 5d supersymmetric higher spin gauge theory.

We consider both unreduced model based on $cu(1,1|8)$ with all
fields appearing in infinitely many copies and the reduced model
based on the superalgebra $hu_0(1,1|8)$, in which every
supermultiplet appears just once. For these particular models we
build  higher spin actions that describe  properly, both at the
free field level and at the level of  cubic interactions, the
systems of totally symmetric boson and fermion $5d$ higher spin
gauge fields with spins $s\geq 3/2$, interacting with gravity. Let
us note that the constructed higher-spin cubic vertices do not
necessarily exhaust all possible interactions in the order under
consideration. The full structure of the cubic action can only be
fixed from the analysis of higher orders, which problem is beyond
the scope of this paper.

Let us  note  that, our formulation operates in terms of
appropriate auxiliary and extra fields identified with particular
higher spin connections. These auxiliary variables simplify the
formulation enormously, being expressed in terms of derivatives of
the particular physical higher spin fields (modulo pure gauge
degrees of freedom) by virtue of appropriate constraints
\cite{LV,vf}. This is analogous to the formulation of gravity by
requiring the metric postulate to be true to define connection in
terms of derivatives of the metric tensor instead of rewriting the
Einstein action directly in terms of the metric tensor. In this
paper we impose the generalized ``higher spin metric postulate''
constraint conditions. The explicit expressions for the auxiliary
fields in terms of the physical ones are not discussed here
because, as is clear from the corresponding $4d$ analysis of
\cite{V1}, the particular expressions are not very illuminating.
It is however straightforward to figure out a form of some
particular cubic vertex for pysical fields by solving appropriate
constraints which have a form of linear algebraic equations on the
auxiliary variables (see section 3).

As argued in \cite{VD5}, it is not straightforward to incorporate
an extended supersymmetry with $\cal{N}\geq $ $2$  in the present
construction of cubic higher spin couplings. This is because
$\cal{N}\geq $ $2$ supermultiplets originated from $cu(2^{{\cal
N}-1}, 2^{{\cal N}-1}|8 )$ require mixed symmetry higher spin
fields in $AdS_5$ to be included\footnote{To avoid
misunderstandings, let us note that what we call ${\cal N}$
extended $AdS_5$ supersymmetry in this paper in some other works
(see e.g. \cite{Gun} and references therein) is referred to as
$2{\cal  N}$ extended $AdS_5$ supersymmetry.}. However, Lagrangian
formulation of such fields in $AdS$ spacetime, is not yet
elaborated in full details even at the free field level, although
a significant progress was achieved recently in
\cite{BMV,Metsaev,BS} \footnote{The situation with the equations
of motion for mixed-symmetry higher spin fields is simpler. The
gauge invariant equations of motion for all types of massless
fields in $AdS_d$ for even $d$ were found in \cite{Metsaev:re}.
Lorentz covariant equations of motion for some particular higher
spin fields in $AdS_5$ with special values of energy $E_0$ were
studied in \cite{SS2}.}.

The paper is organized as follows. In section \ref{5d Higher Spin
Algebra} we recall the construction of ${\cal N} =1$ $AdS_5$
higher spin superalgebra $cu(1,1|8)$ in terms of star product
algebras of superoscillators and define appropriate reality
conditions. Gauging of $cu(1,1|8)$ is studied in section \ref{d5
Higher Spin Gauge Fields}. The construction of the $AdS_5$ higher
spin action functional is the content of section \ref{5d Higher
Spin Action} where, at first, in section 4.1  we discuss general
properties of the higher spin action and give the final output of
our analysis, and then explicitly derive the quadratic (section
\ref{Quadratic Action}) and cubic (section \ref{unred}) higher
spin actions possessing necessary higher spin symmetries.
Reduction to a higher spin gauge theory associated with the
reduced algebra $hu_0(1,1|8)$, in which every integer spin
supermultiplet appears in one copy, is performed  in section
\ref{redmod}. Section \ref{Conclusion} contains conclusions. Some
technicalities are collected in two Appendices.

\section{$5d$ Higher Spin Superalgebra}
\label{5d Higher Spin Algebra}

Consider the associative Weyl-Clifford algebra with
(anti)commutation generating relations \be \label{spa} \ba{l}
[a_{\alpha}, b^{\beta}]_\star=\delta_{\alpha}{}^{\beta}\;,\qquad
[a_{\alpha}, a_{\beta}]_\star=[b^{\alpha},
b^{\beta}]_\star=0\;,\quad \alpha,\beta=1,...,4\;,
\\
\\
\{\psi, \bpsi\}_\star=1\;,\qquad \{\psi, \psi\}_\star
=\{\bpsi,\bpsi\}_\star=0\; \ea \ee with respect to Weyl star
product \be \label{sp} (F\star G)(a,b,\psi,\bpsi)=
F(a,b,\psi,\bpsi)\:(\exp\triangle)\:G(a,b,\psi,\bpsi)\,, \ee where
\be \triangle = \frac{1}{2}\left(\frac{\overleftarrow{\d}}{\d
a_\alpha}\: \frac{\overrightarrow{\d}}{\d b^\alpha}-
\frac{\overleftarrow{\d}}{\d b^\alpha}\:
\frac{\overrightarrow{\d}}{\d a_\alpha} +
\frac{\overleftarrow{\d}}{\d \psi}\: \frac{\overrightarrow{\d}}{\d
\bpsi} +\frac{\overleftarrow{\d}}{\d \bpsi}\:
\frac{\overrightarrow{\d}}{\d \psi}\right). \ee The generators \be
\label{gl} \ba{l} \dps T_\alpha{}^\beta = a_\alpha b^\beta \equiv
\half (a_\alpha\star b^\beta +b^\beta\star a_\alpha)\;,
\\
\\
Q_\alpha=a_\alpha\bpsi\;, \qquad \bar{Q}^\beta= b^\beta\psi\;,
\\
\\
\dps U=\psi\bpsi\equiv \half(\psi\star\bpsi -\bpsi\star\psi)\;
\\
\\
\ea \ee close to the superalgebra $gl(4|1;{\bf C})$ with respect
to the graded Lie supercommutator \be \label{[]*} [F\,,G\}_\star
=F\star G - (-1)^{\pi(F)\pi(G)}G\star F \,, \ee where the $Z_2$
grading $\pi$ is defined by \be
F(-a,-b,\psi,\bpsi)=(-1)^{\pi(F)}F(a,b,\psi,\bpsi)\,,\qquad
\pi(F)=\mbox{0 or 1}. \ee The set of generators (\ref{gl})
consists of $gl(4;{\bf C})$ generators $T$, supersymmetry
generators $Q$ and $\bar{Q}$ and $u(1)$ generator $U$. The central
element in $gl(4|1;{\bf C})$ is \be \label{centrN} N = a_\alpha
b^\alpha -\psi\bpsi\;. \ee The generators of $sl(4|1;{\bf C})$ are
\be \label{str_sl} \dps t_\alpha{}^\beta = a_\alpha
b^\beta-\delta_\alpha{}^\beta\,\psi\bpsi \,,\quad
q_\alpha=a_\alpha\bpsi\;, \quad \bar{q}^\beta= b^\beta \psi\,. \ee
The $\ads$ superalgebra $su(2,2|1)$ \cite{nahm} is a real form of
$sl(4|1;{\bf C})$ singled out by the reality conditions defined
below.

A natural higher spin extension of $su(2,2|1)$ introduced in
\cite{FLA} under the name $shsc^{\infty}(4|1)$ and called
$cu(1,1|8)$ in \cite{Vc}\footnote{The reason for introducing a new
name $cu(1,1|8)$ in \cite{Vc} was to make it possible to include
this particular algebra into the infinite set of algebras
$cu(n,m|2k)$ with different inner symmetries (i.e., Chan-Paton
factors) labelled by two non-negative integers $n,m$, as well as
to allow an arbitrary number of indices $\alpha,\beta= 1,...,k.$}
is associated with the star product algebra of all polynomials
$F(a,b,\psi,\bpsi)$ satisfying the condition \be
\label{dpoc}[N,F]_\star=0\;. \ee In other words, the $5d$ higher
spin superalgebra $cu(1,1|8)$ is spanned by
star-(anti)com\-mutators of the elements of the centralizer of $N$
in the star product algebra (\ref{spa}). As a corollary, every $F$
satisfying (\ref{dpoc}) has the form \be \label{basis} \ba{c} \dps
F(a,b,\psi,\bpsi) \equiv A(a,b)+B(a,b)\psi+D(a,b)\bpsi
+E(a,b)\psi\bpsi
\\
\\
\dps
=\sum_{k=0}^{\infty}A^{\alpha(k)}_{\:\beta(k)}\:a_{\alpha(k)}\:b^{\beta(k)}
\dps
+\sum_{k=0}^{\infty}B^{\alpha(k)}_{\:\beta(k+1)}\:a_{\alpha(k)}\:b^{\beta(k+1)}\;\psi
\\
\\
\dps
+\sum_{k=0}^{\infty}D^{\alpha(k+1)}_{\:\beta(k)}\:a_{\alpha(k+1)}\:b^{\beta(k)}\;\bpsi
\dps +
\sum_{k=0}^{\infty}E^{\alpha(k)}_{\:\beta(k)}\:a_{\alpha(k)}\:b^{\beta(k)}\;\psi\bpsi\;,
\ea \ee where we use notations \be a_{\alpha(k)}\equiv
a_{\alpha_1} \ldots a_{\alpha_k}\;, \qquad b^{\beta(k)}\equiv
b^{\beta_1} \dots b^{\beta_k} \ee and
$A^{\alpha(k)}_{\:\beta(k)},B^{\alpha(k)}_{\:\beta(k+1)},D^{\alpha(k+1)}_{\:\beta(k)}$
and $E^{\alpha(k)}_{\:\beta(k)}$ are arbitrary multispinors
totally symmetric in lower and upper indices\footnote{When
handling multispinors we adhere conventions introduced in
\cite{V1}. Namely, a number of symmetrized indices is indicated in
parentheses.  Lower and upper indices denoted by the same letter
are separately symmetrized and then a maximal possible number of
lower and upper indices denoted by the same letter has to be
contracted.}. Note that $F\in cu(1,1|8)$ is  even in
superoscillators.

To single out an appropriate real form of the complex higher spin
algebra $cu(1,1|8)$ we impose reality conditions in the following
way. Introduce an involution $\dagger$ defined by the relations
\be \label{inv} ({a}_\alpha )^\dagger  = i b^\beta C_{\beta
\alpha}\,,\qquad ({b}^\alpha )^\dagger = i C^{\alpha\beta} a_\beta
\,, \ee \be \label{invf} (\psi)^\dagger = \bpsi\;,\qquad
(\bpsi)^\dagger = \psi\;, \ee where
$C_{\alpha\beta}=-C_{\beta\alpha}$ and
$C^{\alpha\beta}=-C^{\beta\alpha}$ are some real antisymmetric
matrices satisfying \be C_{\alpha\gamma} C^{\beta\gamma} = \delta
_\alpha^\beta\,. \ee An involution is required to reverse an order
of product factors \be \label{ord} (F\star G)^\dagger = G^\dagger
\star F^\dagger\, \ee and to conjugate complex numbers \be
\label{alin} (\mu F)^\dagger = \bar{\mu} F^\dagger\,,\qquad \mu
\in {\bf C}\,, \ee where the bar denotes complex conjugation. The
involution $\dagger$ leaves invariant the defining relations
(\ref{spa}) of the star product algebra and satisfies $(\dagger
)^2 = Id$. By (\ref{ord}) the action (\ref{inv}), (\ref{invf}) of
$\dagger$ extends to an arbitrary element $F$ of the star product
algebra. Since the star product we use corresponds to the totally
(anti)symmetric (i.e. Weyl) ordering of the product factors, the
result is \be \label{dag} \ba{c} (F(a_\alpha ,b^\beta,\psi,\bpsi
))^\dagger = \bar{A}(i b^\gamma C_{\gamma \alpha} ,i
C^{\beta\gamma} a_\gamma)+ \bar{D}(i b^\gamma C_{\gamma \alpha}
,iC^{\beta\gamma} a_\gamma) \psi
\\
\\
+\bar{B}(i b^\gamma C_{\gamma \alpha} ,i C^{\beta\gamma}
a_\gamma)\bpsi + \bar{E}(i b^\gamma C_{\gamma \alpha} ,i
C^{\beta\gamma} a_\gamma)\psi\bpsi \,. \ea \ee The  involution
$\dagger$ (\ref{dag}) allows us to define a real form of the Lie
superalgebra built by virtue of graded commutators of elements
(\ref{basis}) by imposing the condition (for more details see e.g.
\cite{Fort2}) \be \label{reco} F^\dagger = -i^{\pi(F)}  F\,. \ee
This condition defines the real higher spin algebra $cu(1,1|8)$
\cite{Vc}. It contains the ${\cal N}=1$ $AdS_5$ superalgebra
$su(2,2|1)$ as its finite-dimensional subalgebra. In fact, the
reality condition (\ref{reco}) guarantees that $cu(1,1|8)$  admits
massless unitary representations with energy bounded below
\cite{VK}.

\section{$5d$ Higher Spin Gauge Fields}
\label{d5 Higher Spin Gauge Fields}

The $\ads$ totally symmetric higher spin gauge fields can be
described \cite{LV,vf,SS1,VD5,A1,SS2} in terms of 1-form gauge
fields $\Omega(a, b, \psi, \bpsi
|x)=dx^{\underline{n}}\Omega_{\underline{n}}(a, b, \psi, \bpsi
|x)$ ($\underline{n}=0,...,4$) of $cu(1,1|8)$ \be \label{gf}
\Omega(a, b, \psi, \bpsi |x) =
\Omega{_{E_1}}(a,b|x)+\Omega{_{O_1}}(a,b|x)\psi +
\Omega_{O_2}(a,b|x)\bpsi +\Omega_{E_2}(a,b|x) \psi\bpsi\,, \ee
where \be \label{gf1}
\Omega_{E_1}(a,b|x)=\sum_{k=0}^{\infty}(\Omega_{E_1}(x))^{\alpha(k)}_{\beta(k)}a_{\alpha(k)}b^{\beta(k)}\;,
\ee \be \label{gf2}
\Omega_{E_2}(a,b|x)=\sum_{k=0}^{\infty}(\Omega_{E_2}(x))^{\alpha(k)}_{\beta(k)}a_{\alpha(k)}b^{\beta(k)}\;
\ee with commuting multispinors
$(\Omega_{E_{1,2}}(x))^{\alpha(m)}_{\beta(m)}$ (label E means
"even") and \be \label{gf3}
\Omega_{O_1}(a,b|x)=\sum_{k=0}^{\infty}(\Omega_{O_1}(x))^{\alpha(k)}_{\beta(k+1)}a_{\alpha(k)}b^{\beta(k+1)}\;,
\ee \be \label{gf4}
\Omega_{O_2}(a,b|x)=\sum_{k=0}^{\infty}(\Omega_{O_2}(x))^{\alpha(k+1)}_{\beta(k)}a_{\alpha(k+1)}b^{\beta(k)}\;
\ee with anticommuting multispinors
$(\Omega_{O_{1,2}}(x))^{\alpha(m)}_{\beta(n)},\,|m-n|=1$ (label O
means "odd"). We require the component gauge fields
$\Omega^{\alpha(m)}_{\beta(n)}(x)=
dx^{\underline{n}}\Omega{}^{\;\;\;\alpha(m)}_{\underline{n}\,\;\beta(n)}(x),\,|m-n|\leq
1$ to commute with the basis elements of $cu(1,1|8)$ (i.e. with
the superoscillators $a_\alpha, b^\beta, \psi$ and $\bpsi$).

The higher spin field strength $ R(a, b, \psi, \bpsi |x)\equiv R$
\be \label{Rosn} R= d\Omega + \Omega  \wedge\star\,
\Omega\,,\qquad d=dx^{\underline{n}}\frac{\d}{\d
x^{\underline{n}}} \ee admits an expansion analogous to
(\ref{gf})-(\ref{gf4}). Infinitesimal higher spin gauge
transformations are \be \label{gotr} \delta\Omega = D\epsilon\,,
\qquad \delta R =  [R  \,, \epsilon ]_\star\,, \ee where 0-form
$\epsilon=\epsilon(a,b,\psi,\bar{\psi}|x)$ is an arbitrary
infinitesimal higher spin gauge symmetry parameter and \be
\label{Dc} D F = d F + [\Omega  \,, F ]_\star\,. \ee

To analyse interactions we will use the perturbation expansion
with the dynamical fields $\Omega_1$ treated as fluctuations above
the appropriately chosen background $\Omega_0$ \be \label{go01}
\Omega = \Omega_0 +\Omega_1 \,, \ee where the vacuum gauge fields
$\Omega_0=\Omega_{0\,\beta}^{\;\;\alpha}(x)\, a_\alpha b^\beta$
correspond to background $AdS_5$ geometry described by virtue of
the zero-curvature condition $R(\Omega_0 )\equiv d\Omega_0+
\Omega_0 \wedge\star\, \Omega_0=0$ (for more details see Appendix
A of this paper and \cite{V_obz2,VD5}). Since $R(\Omega_0)=0$, we
have $R=R_1 +R_2\,,$ where \be \label{R11} R_1 = d\Omega_1
+\Omega_0 \star\wedge \Omega_1 +  \Omega_1\star\wedge
\Omega_0\,,\qquad R_2 = \Omega_1 \star \wedge  \Omega_1\,. \ee The
Abelian lowest order part of the transformation (\ref{gotr}) has
the form \be \label{lintr} \delta_0 \Omega_1 =  D_0 \epsilon
\,,\qquad \delta_0 R_1 = 0\, \ee with the covariant derivative
$D_0$ (\ref{Dc}) evaluated with respect to the background field
$\Omega_0$.

The higher spin gauge fields of the real higher spin algebra
$cu(1,1|8)$ singled out by the conditions (\ref{reco}), satisfy
the reality conditions \cite{Fort2,Vc} \be \label{rego}
\Omega^\dag = -i^{\pi(\Omega)} \Omega\,. \ee In fact, this
condition implies that the odd component fields
$(\Omega_{O1})^{\alpha(s)}_{\beta(s+1)}(x)$ and
\newline
$(\Omega_{O_2})^{\alpha(s+1)}_{\beta(s)}(x)$  are conjugated to
each other while the even component fields
$(\Omega_{E_{1,2}})^{\alpha(s)}_{\beta(s)}(x)$ are
self-conjugated.

In accordance with the analysis of \cite{LV,vf,SS1,VD5,A1,SS2}
$5d$ totally symmetric higher spin fields can be described by
1-forms $\Omega^{\alpha(m)}_{\beta(n)}(x)\equiv
dx^{\underline{n}}\Omega_{\underline{n}}{}^{\alpha(m)}_{\:\beta(n)}(x)\,,|m-n|\leq
1\,,$ being traceless multispinors symmetric separately in the
upper and lower indices. The case of $m=n=s$
 corresponds to the bosonic spin
$s^\prime=s+1$ field while the cases of $n=s$, $m=s+1$ and
$n=s+1$, $m=s$ correspond to the fermionic spin $s^\prime=s+3/2$
field. Thus, even and odd multispinors in (\ref{gf1})-(\ref{gf4})
are identified with bosonic and fermionic totally symmetric higher
spin fields, respectively. As shown in \cite{vf}, the number of
on-shell degrees of freedom $deg(m,n)$ described  by
$\Omega_{\underline{n}}{}^{\alpha(m)}_{\:\beta(n)}(x)\,,|m-n|\leq
1\,,$ is given by \be \label{degreeB} deg(m,n)= \left\{ \ba{l}
2s+3\,,\;\;\;\;n=m=s\,, \\ 4(s+2)\,,\;m=s+1\, {\rm or}
\;\;n=s+1\,, \ea \right. \ee being  precisely the (real)
dimensionalities of the corresponding (spin)-tensor irreps of the
little group $SO(3)$. The multiplet $(s,s-\half, s-1)$ therefore
contains equal numbers of boson and fermion degrees of freedom.

The multispinors in (\ref{gf1})-(\ref{gf4}) are not traceless and,
therefore, each of them decomposes into a sum of irreducible
traceless components. Namely, for any fixed $n$ and $m$, tensor
$\Omega^{\alpha(m)}_{\beta(n)}(x)$ decomposes into the set of
irreducible traceless components
$\Omega^{\prime\,\alpha(k)}_{\;\beta(l)}(x)\,,$
($\Omega^{\prime\,\alpha(k-1)\gamma}_{\;\beta(l-1)\gamma}(x) =0$)
with all $k+l\leq n+m$, $k-l = n-m$, $k\geq 0$, $l\geq 0$. As a
result,  a field of every spin appears in infinitely many copies
in the expansion (\ref{gf})-(\ref{gf4}) \be \label{Vs} \Omega =
\sum_{k=0}^{\infty}\;\sum_{s=2}^{\infty}\;D^{(k)}(s)\oplus
D^{(k)}(s-\frac{1}{2}) \oplus\bar{D}^{(k)}(s-\frac{1}{2})\oplus
D^{(k)}(s-1)\;, \ee where $D^{(k)}(s)$ denotes a k-th copy of spin
$s$ $su(2,2)$ irreducible representation carried by traceless
multispinors in the 1-form $\Omega^{\alpha(m)}_{\beta(n)}$,
$|m-n|\leq 1$.

The origin of this infinite degeneracy can be traced back to the
fact that the algebra $cu(1,1|8)$ is not simple but contains
infinitely many ideals $I_{P(N)}$, where $P(N)$ is any
star-polynomial of $N$, spanned by the elements of the form
$\{x\in I_{P(N)} : x=P(N)\star F,\;\; F\in cu(1,1|8)\}$
\cite{FLA}. One may consider quotient algebras
$cu(1,1|8)/I_{P(N)}$. The most interesting reduction is provided
by the algebra $hu_0(1,1|8)=cu(1,1|8)\!/I_N$, where $I_N$ is the
ideal spanned by the elements $x= N\star F = F\star N$. The higher
spin model with spectra of spins associated with $hu_0(1,1|8)$ is
built in section \ref{redmod}.

For the future convenience we introduce the two sets of the
differential operators in the auxiliary variables \be \label{T}
T^+ = a_\alpha b^\alpha\,,\qquad T^- =\frac{1}{4}\frac{\d^2 }{ \d
a_\alpha \d b^\alpha } \,,\qquad T^0 =\frac{1}{4} ( N_a +N_b  +4
)\; \ee and \be \label{NN} P^+ = T^+-\psi\bpsi\,,\quad P^- =T^-
+\frac{1}{4}\frac{\d^2 }{ \d \bpsi \d \psi } \,, \quad P^0
=T^0+\frac{1}{4}(N_{\psi} +N_{\bpsi}-1)\,\,, \ee where \be \ba{cc}
\label{N} \dps N_a=a_\alpha\frac{\d}{\d a_\alpha}\,, & \dps
N_b=b^\alpha\frac{\d}{\d b^\alpha}\,,
\\ & \\
\dps N_{\psi}=\psi\frac{\d}{\d \psi}\,, & \dps
N_{\bpsi}=\bpsi\frac{\d}{\d \bpsi}\,.

\ea \ee These operators form the $sl_2$ algebras \be
\label{sl2inv} [ T^0 , T^\pm ] =  \pm  \frac{1}{2} T^\pm \,,\qquad
[T^- , T^+ ] = T^0\,, \ee \be \label{slN} [ P^0 , P^\pm] =  \pm
\frac{1}{2} P^\pm \,,\qquad [P^- , P^+ ] =  P^0\,. \ee

Expansion coefficients of an element $\Omega(a,b,\psi,\bpsi|x)$
are supertraceless iff \\ $ P^-\, \Omega(a,b,\psi,\bpsi|x) =0\,.$
As a result, the operators $P^-$ and $P^+$ allow one to write down
the decomposition of an arbitrary element
$\Omega(a,b,\psi,\bpsi|x)$ of $cu(1,1|8)$ into irreducible
$su(2,2|1)$ supermultiplets as \be \label{STRD}
\Omega(a,b,\psi,\bpsi|x) =
\sum_{k=0}^{\infty}\,\sum_{s=1}^{\infty}\,\chi(k,\,s)\,(P^+)^k
\;\Omega^{k,\,s+1}(a,b,\psi,\bpsi|x)\,, \ee where $\chi(k,\,s)$
are some non-zero normalization coefficients, $s+1$ denotes
highest integer spin in a supermultiplet and $\Omega^{k,s+1}$
defined by $P^0\Omega^{k,\,s+1} =(2s+3)/4\,\Omega^{k,\,s+1}$ are
supertraceless \be \label{STR}
P^-\,\Omega^{k,\,s+1}(a,b,\psi,\bpsi|x) =0\,. \ee The condition
(\ref{STR}) solves explicitly as \be \label{AB1} \ba{c} \dps
\Omega^{k,s+1}(a, b, \psi, \bpsi
|x)=\tilde{\Omega}^{k,s+1}_{E_1}(a,b|x) -\frac{1}{(2s+2)}T^+
\tilde{\Omega}^{k,s}_{E_2}(a,b|x)
\\
\\
+\tilde{\Omega}^{k,s+\frac{1}{2}}_{O_1}(a,b|x)\psi
+\tilde{\Omega}^{k,s+\frac{1}{2}}_{O_2}(a,b|x)\bpsi
+\tilde{\Omega}^{k,s}_{E_2}(a,b|x)\psi\bpsi\;,

\ea \ee where all $su(2,2)$ multispinors are traceless \be
\label{AB2}
T^-\tilde{\Omega}^{k,s^\prime}_{E_{1,2}}(a,b|x)=T^-\tilde{\Omega}^{k,s^\prime}_{O_{1,2}}(a,b|x)=0\;,
\quad s^\prime = s,s+\frac{1}{2},s+1\;. \ee Thus, the gauge fields
originating from $cu(1,1|8)$ are arranged into an infinite
sequence of supermultiplets $\{s^{\prime}\}^{(k)}, 0\leq k
<\infty$, with a spin content
$(s^{\prime},s^{\prime}-\frac{1}{2},s^{\prime}-1)^{(k)}$
determined by an integer highest spin $s^{\prime}=2,3,...,\infty
\;$.

The decomposition (\ref{STRD}) can be rewritten in the $su(2,2)$
basis with all multispinors being traceless rather than
supertraceless. The two bases are related by a finite field
redefinition. The final result derived in Appendix B is \be
\label{tr1} \Omega_{E_{1,2}}(a,b|x) = \sum_{n,\,s=0}^{\infty} \,
v_{E_{1,2},n}(T^0)\;(T^+)^n \;\Omega_{E_{1,2}}^{n,s+1}(a,b|x)\;,
\ee \be \label{tr2} \Omega_{O_{1,2}}(a,b|x) =
\sum_{n,\,s=0}^{\infty} \, v_{O_{1,2},n} (T^0)\;(T^+)^n
\;\Omega_{O_{1,2}}^{n,s+3/2}(a,b|x)\,, \ee where $v_{E_{1,2},n}$
and $v_{O_{1,2},n}$ are some  non-zero normalization  coefficients
and \be \label{irrr1} T^0\,\Omega_{E_{1,2}}^{n,s+1}(a,b|x)
=\frac{1}{2}(s+2)\;\Omega_{E_{1,2}}^{n,s+1}(a,b|x)\;, \ee

\be \label{irrr3} T^0\,\Omega_{O_{1,2}}^{n,s+3/2}(a,b|x)=
\frac{1}{4}(2s+5)\;\Omega_{O_{1,2}}^{n,s+3/2}(a,b|x)\;, \ee

\be \label{7okt.} T^-\,\Omega_{E_{1,2}}^{n,s+1}(a,b|x) =
T^-\,\Omega_{O_{1,2}}^{n,s+3/2}(a,b|x) = 0\;. \ee For the future
convenience, we fix the normalization coefficients in the form \be
\label{coe}
v_{E_{1,2},n}(s)=(2i)^n\;\sqrt{\frac{(2s+3)!}{n!(2s+3+n)!}}\;\;,
\ee

\be \label{coe1}
v_{O_{1,2},n}(s)=(2i)^n\;\sqrt{\frac{(2s+4)!}{n!(2s+4+n)!}}\;\;,
\ee where the factor of $i^n$ is introduced because the operator
$T^+$ is antihermitian.

It is worth noting that unlike the supertrace decomposition
(\ref{STRD}), the fields carrying the same label $n$ in
(\ref{tr1})-(\ref{tr2}) may belong to different supermultiplets.

In addition to (\ref{7okt.}) fields $\Omega(a,b|x)$ satisfy the
conditions \be \label{ir1} \ba{c} (1+N_a-N_b)\Omega_{O_1}(a,b|x)
=0\;, \quad (1+N_b-N_a)\Omega_{O_2}(a,b|x) =0\;,
\\
\\
(N_b-N_a)\Omega_{E_{1,2}}(a,b|x) =0\;, \ea \ee that express the
condition (\ref{dpoc}).

The operators $T^i$ (\ref{T}) are $su(2,2)$ invariant. As a result
\be \label{DT0} D_0 (T^i ) =0\,, \ee which relations have to be
understood in the sense that $D_0 (X(F)) = X(D_0 (F))$, where $X$
is one of the operators $T^i$, while $F$ is an arbitrary element
of the higher spin algebra. A useful consequence of this fact is
\be \label{RT} R_1 (T^j (\Omega)) = T^j (R_1 ( \Omega ) )\,, \ee
where $R_1$ denotes the linearised higher spin curvature
(\ref{R11}). Due to (\ref{RT}), the linearised curvatures admit
the expansion analogous to (\ref{tr1})-(\ref{tr2}): \be
\label{tr1_R} R_{1,\,E_{1,2}}(a,b|x) = \sum_{n,\,s=0}^{\infty} \,
v_{E_{1,2},n}(T^0) \;(T^+)^n \;R_{1,\,E_{1,2}}^{n,s+1}(a,b|x)\;,
\ee \be \label{tr2_R} R_{1,\,O_{1,2}}(a,b|x) =
\sum_{n,\,s=0}^{\infty} \, v_{O_{1,2},n} (T^0)\;(T^+)^n
\;R_{1,\,O_{1,2}}^{n,s+3/2}(a,b|x)\,, \ee where the curvatures on
r.h.s.'s satisfy the irreducibility conditions analogous to
(\ref{irrr1})-(\ref{ir1}).

The $su(2,2)$ irreducible higher spin gauge field
$\Omega^{n,s^\prime}$ decomposes into a set of Lorentz covariant
fields that form irreducible representations of the Lor\-entz
algebra $so(4,1)\subset su(2,2)$. Different Lorentz gauge fields
get different dynamical interpretation. For example, the $su(2,2)$
irreducible field $\Omega^\alpha_\beta(x)$ in the adjoint of
$su(2,2)$ used to describe spin 2 field contains the frame field
and Lorentz connection as the different irreducible Lorentz
components. To decompose $su(2,2)$ representations into Lorentz
irreps we make use of the compensator formalism. (For more details
on the compensator formalism and the decomposition procedure see
\cite{VD5} and Appendix A of this paper.) Namely, let
$V^{\alpha\beta}=-V^{\beta\alpha}$ be a nondegenerate
antisymmetric matrix. Then, the Lorentz subalgebra of $su(2,2)$
can be defined as the stability algebra of $V^{\alpha\beta}$. In
fact, this can be done locally with $V^{\alpha\beta}(x)$ being a
field. We shall treat $V^{\alpha\beta}$ as a symplectic form that
allows to raise and lower spinor indices in the  Lorentz covariant
way

\be A^\alpha = V^{\alpha\beta}A_\beta \,,\qquad  A_\alpha =
A^\beta V_{\beta\alpha}\;. \ee Let us introduce the operators \be
\label{S} S^- =V^{\alpha\beta} a_\alpha \frac{\d}{\d b^\beta}
\,,\qquad S^+ =V_{\alpha\beta} b^\alpha \frac{\d}{\d a_\beta}
\,,\qquad S^0 = N_b - N_a\;, \ee satisfying the commutations
relations \be \label{sl2S} [ S^0 , S^\pm ] =  \pm  2 S^\pm
\,,\qquad [S^- , S^+ ] = S^0\,,\qquad [S^i, T^j]=0\,. \ee With the
help of the operators (\ref{S}) the decomposition into the higher
spin Lorentz irreducible 1-forms is given by \be \label{expan1}
\Omega_{E_{1,2}}(a,b|x) = \sum_{t=0}^{s} (S^+)^t
\:\omega_{e_{1,2}}^t(a,b|x), \ee where \be \label{expan2}
\omega_{e_{1,2}}^t(a,b|x) = \omega_{e_{1,2}}^{\alpha(s+t),\;
\beta(s-t)}\,(x)\; a_{\alpha(s+t)} b_{\beta(s-t)} \ee are bosonic
fields and \be \label{expan3} \Omega_{O_{1}}(a,b|x) =
\sum_{t=0}^{s} (S^-)^t \:\omega_{o_{1}}^t(a,b|x)\;, \ee \be
\label{expan3per} \Omega_{O_{2}}(a,b|x) = \sum_{t=0}^{s} (S^+)^t
\:\omega_{o_{2}}^t(a,b|x)\;, \ee where \be \label{expan4}
\omega_{o_{1}}^t(a,b|x) =
\omega_{o_{1}}^{\beta(s+t+1),\;\alpha(s-t) }\,(x)\;
a_{\alpha(s-t)} b_{\beta(s+t+1)}\;, \ee \be \label{expan4ker}
\omega_{o_{2}}^t(a,b|x) =
\omega_{o_{2}}^{\alpha(s+t+1),\;\beta(s-t) }\,(x)\;
a_{\alpha(s+t+1)} b_{\beta(s-t)}\; \ee are fermionic fields. With
respect to their tangent indices Lorentz higher spin fields
$\omega_{e_{1,2}}$ (\ref{expan2}) and $\omega_{o_{1,2}}$
(\ref{expan4}), (\ref{expan4ker}) are described by the traceless
two-row Young diagrams, i.e. \be \label{irr} \ba{c}
S^-\omega_{e_{1,2}}^t(a,b|x) = 0\;, \qquad
S^-\omega_{o_2}^t(a,b|x) = 0 \;,\qquad S^+\omega_{o_1}^t(a,b|x) =
0\;,
\\
\\
T^-\omega_{e_{1,2}}^t(a,b|x) = 0\;,\qquad
T^-\omega_{o_{1,2}}^t(a,b|x) = 0\;. \ea \ee The Lorentz higher
spin curvatures $r^t(a,b|x)$ associated with the fields
$\omega^t(a,b|x)$ are defined by means of analogous procedure
applied to $R(a,b|x)$. Their form in terms of Lorentz gauge fields
is as follows \be \label{taus1} r_e^t = {\cal D} \omega_e^t
+\tau^-\omega_e^{t+1}+\tau^+\omega_e^{t-1}\;, \ee \be
\label{taus2} r_o^t = {\cal D} \omega_o^t  +{\cal
T}^-\omega_o^{t+1} +{\cal T}^0\omega_o^t +{\cal
T}^+\omega_o^{t-1}\;, \ee where ${\cal D}$ is the background
Lorentz derivative. The explicit expressions for the operators
$\tau$ and ${\cal T}$ are given in \cite{VD5} and \cite{A1},
respectively. The corresponding gauge transformation laws
 (\ref{lintr})  have the form analogous to
(\ref{taus1})-(\ref{taus2}).

{}From the dynamical point of view bosonic $\omega_e$  and
fermionic $\omega_o$ fields (\ref{expan2}), (\ref{expan4}),
(\ref{expan4ker}) with $t=0$ are analogous to the frame field and
gravitino and are treated as dynamical fields $\omega^{ph}$ while
all other fields with $t>0$ play a role analogous to Lorentz
connection. These are either auxiliary fields ($t=1$ for bosons)
or ``extra" fields ($t\geq 2$ for bosons and $t\geq 1$ for
fermions). Extra fields do not contribute into the free action
functional since its variation  w.r.t. extra fields is required to
be zero identically. (This is the so called {\it extra field
decoupling condition}\,; see section \ref{5d Higher Spin Action}.)
However, these fields do contribute at the interaction level. To
make such interactions meaningful, one has to express the
auxiliary and extra fields in terms of the physical ones modulo
pure gauge degrees of freedom. This is achieved by imposing
appropriately chosen constraints \cite{LV,vf} which have the form
\be \label{constraints} \Upsilon^+_2\wedge r^t_1 = 0\;,\quad 0\leq
t < s\;, \ee where  $r_1^t$ are Lorentz linearized curvatures
(\ref{taus1})-(\ref{taus2}) and \be \label{upsilon} \Upsilon^+_2=
\left\{ \ba{l} \tau^0\wedge \tau^{+} \,,\;\;\;\;{\rm for
\;\;bosons}\,,
\\

{\cal T}^0\wedge {\cal T}^{+}\,,\;{\rm \;for \;\;fermions}\,. \ea
\right. \ee is such  a 2-form operator that the number of
independent algebraic conditions, imposed on the curvature
components $r_1^t$ by (\ref{constraints}) coincides with the
number of components of the extra field $\omega^{t+1}$ minus the
number of its pure gauge components. For  explicit expressions of
the tau operators we refer the reader to
 \cite{VD5,A1}.

An important fact is that, by virtue of these constraints most of
the higher spin curvatures $r^t(a,b|x)$ vanish on mass-shell
according to the following relationship referred to as the First
On-Mass-Shell Theorem \cite{LV,vf,SS1,VD5,SS2}: \be \label{2nov}
\ba{l} \dps r^{\alpha(s+t),\; \beta(s-t)}(x)\: =X^{\alpha(s+t),\;
\beta(s-t)}\Big(\frac{\delta {\cal S}_2}{\delta
\omega_e^{ph}}\Big)\,,\qquad {\rm for}\;\;t<s\,,
\\
\\
\dps r^{\alpha(2s)}(x)\: = h_{\alpha\gamma}\wedge
h^\gamma{}_\alpha
C_e^{\alpha(2s+2)}(x)+X^{\alpha(2s)}\Big(\frac{\delta {\cal
S}_2}{\delta \omega_e^{ph}}\Big)\,,\qquad {\rm for}\;\;t=s\,

\ea \ee for bosons and \be \label{1nov} \ba{c} \dps
r^{\alpha(s+t+1),\; \beta(s-t)}(x)\: =Y^{\alpha(s+t+1),\;
\beta(s-t)}\Big(\frac{\delta {\cal S}_2}{\delta
\omega_o^{ph}}\Big)\,,\qquad {\rm for}\;\;t<s\,,
\\
\\
\dps r^{\alpha(2s+1)}(x)\: = h_{\alpha\gamma}\wedge
h^\gamma{}_\alpha
C_o^{\alpha(2s+3)}(x)+Y^{\alpha(2s+1)}\Big(\frac{\delta {\cal
S}_2}{\delta \omega_o^{ph}}\Big)\,,\qquad {\rm for}\;\;t=s\;,

\ea \ee (plus complex conjugate) for fermions. Here $
h_{\alpha\beta}$ denotes the background frame field (see Appendix
A) and $X$ and $Y$ are some linear functionals of the r.h.s.'s of
the free field equations. The $0$-forms $C_e$ and $C_o$ on the
l.h.s.'s of (\ref{2nov}), (\ref{1nov}) represent generalised Weyl
tensors which are totally symmetric multispinors. The $su(2,2)$
covariant version of (\ref{2nov})-(\ref{1nov}) is \be
\label{onmassO10} {\cal
R}_{E}(a,b|x)\Big|_{m.s.}=H_{2\,\alpha}{}^\beta \frac{\d^2}{\d
a_\alpha \d b^\beta}{\rm Res}_\mu (C_{E}(\mu a+\mu^{-1}b|x))\, \ee
for bosons and \be \label{onmassO11} {\cal
R}_{O_1}(a,b|x)\Big|_{m.s.}=H_{2\,\alpha}{}^\beta \frac{\d^2}{\d
a_\alpha \d b^\beta}{\rm Res}_\mu (\mu \,C_{O_1}(\mu
a+\mu^{-1}b|x))\,, \ee \be \label{onmassO21} {\cal
R}_{O_2}(a,b|x)\Big|_{m.s.}=H_{2\,\alpha}{}^\beta \frac{\d^2}{\d
a_\alpha \d b^\beta}{\rm Res}_\mu (\mu^{-1}\,C_{O_2}(\mu
a+\mu^{-1}b|x))\, \ee for fermions. Here
$H_{2\,\alpha\beta}=h_{\alpha\gamma}\wedge h^\gamma{}_\beta$, the
label $\Big |_{m.s.}$ implies the on-mass-shell consideration
$\frac{\delta {\cal S}_2}{\delta \omega^{ph}}=0$ and $Res_\mu $
singles out the $\mu-$independent part of Laurent series in $\mu$.
Note that a function of one spinor variable \be C(\mu a + \mu^{-1}
b) = \sum_{k,\,l} \frac{\mu^{k-l}}{k!\, l!} C^{\alpha_1 \ldots
\alpha_{k}\beta_1 \ldots \beta_{l}} a_{\alpha_1} \ldots
a_{\alpha_k} b_{\beta_1} \ldots b_{\beta_l} \ee has totally
symmetric coefficients $C^{\alpha_1 \ldots \alpha_{k}\beta_1
\ldots \beta_{l}}$ while $Res_\mu$ in (\ref{onmassO10})-
(\ref{onmassO21}) singles out its part that belongs to $cu(1,1|8)$
with the numbers of the oscillators $a$ and $b$ differing by at
most 1.

\section{${\cal N}=1$ Supersymmetric Higher Spin Action}
\label{5d Higher Spin Action}

The aim of this section is to formulate the action for the $AdS_5$
massless boson and fermion gauge fields of $cu(1,1|8)$ that solves
the problem of higher-spin-gravitational interactions in the first
nontrivial order. The reported results extend the purely bosonic
analysis (${\cal N}=0$) of \cite{VD5} to the ${\cal N}=1$
supersymmetric case.

\subsection{General properties}

The action functional underlying the $5d$ non-linear higher spin
dynamics in the cubic order has the following standard form
\cite{V1,FV1,VD5} \be \label{macmans} {\cal S} = \int U_{12}\wedge
R(\Omega_1)\wedge R(\Omega_2)\,, \ee which is a higher spin
generalization of the MacDowell-Mansouri action for gravity
\cite{MacDowell:1977jt}.
 $U_{12}$ are some 1-form coefficients
built from the frame field and the compensator. $R(\Omega_{1,2})$
are higher spin curvatures associated with higher spin gauge
fields $\Omega_{1,2}$ (\ref{gf}).  Our goal  is to find such
coefficients $U_{12}$ that account for the correct description of
free higher spin dynamics and its consistent non-trivial
interaction deformation. Note that if $U_{12}$ would be a
invariant tensor of the higher spin algebra, the action
(\ref{macmans}) would be a topological invariant thus describing
no non-trivial dynamics. Of course, the main justification of the
form (\ref{macmans}) for the action is that it will be shown to
describe correctly the higher spin dynamics at least in the cubic
order.

Let us now discuss the  structure of the action (\ref{macmans}) in
more detail. An appropriate ansatz is \be \label{acta} {\cal
S}(R,R)=\frac{1}{2}{\cal A}(R,R)\,, \ee where the symmetric
bilinear ${\cal A}(F,G)={\cal A}(G,F)$ is defined for any 2-forms
$F$ and $G$ \be \ba{l} F=F_{E_1}+F_{O_1}\psi+F_{O_2}\bpsi
+F_{E_2}\psi\bpsi\,,
\\
\\
G=G_{E_1}+G_{O_1}\psi+G_{O_2}\bpsi +G_{E_2}\psi\bpsi\,

\ea \ee as \be \label{act} \dps {\cal A}(F,G)= {\cal
B}(F_{E},G_{E})+ {\cal F}(F_{O},G_{O})\;, \ee where \cite{VD5,A1}
\be \label{bosact} \dps {\cal B}(F_{E},G_{E})\equiv{\cal
B}^{\prime}(F_{E_1},G_{E_1}) +{\cal
B}^{\prime\prime}(F_{E_2},G_{E_2})\;, \ee

\be \label{bosact2} \ba{c} \dps{\cal B}^{\prime}(F_{E_1},G_{E_1})=
 \int_{{\cal M}^5} \hat{H}_{E1}\wedge {\rm tr}
(F_{E_1}(a_1,b_1)\wedge G_{E_1}(a_2,b_2))|_{a_i=b_i=0}\,,
\\
\\
\dps {\cal B}^{\prime\prime}(F_{E_2},G_{E_2})= \int_{{\cal M}^5}
\hat{H}_{E2}\wedge {\rm tr} (F_{E_2}(a_1,b_1)\wedge
G_{E_2}(a_2,b_2))|_{a_i=b_i=0}\,,

\ea \ee

\be \label{fermact} \ba{r} \dps {\cal
F}(F_{O},G_{O})=\frac{1}{2}\int_{{\cal M}^5} \hat{H}_{O}\wedge
{\rm tr}(G_{O_2}(a_1,b_1)\wedge F_{O_1}(a_2,b_2))|_{a_i=b_i=0}
\\
\\
\dps +\frac{1}{2}\int_{{\cal M}^5} \hat{H}_{O}\wedge {\rm
tr}(F_{O_2}(a_1,b_1)\wedge G_{O_1}(a_2,b_2))|_{a_i=b_i=0}\,.
\\

\ea \ee 1-forms $\hat{H}_{E1}, \hat{H}_{E2}, \hat{H}_{O}$ are the
following differential operators \be \label{He} \ba{c} \dps
\hat{H}_{E_i}= \alpha_i(p,q,t) E_{\alpha\beta} \frac{\d^2}{\d
a_{1\alpha} \d a_{2\beta}}\hat{b}_{12} +\beta_i(p,q,t)
E^{\alpha\beta} \frac{\d^2}{\d b_1^\alpha \d
b_2^\beta}\hat{a}_{12}
\\
\\
+\dps \gamma_i(p,q,t)(E_\alpha{}^\beta \frac{\d^2}{\d a_{2\alpha}
\d b_1^\beta}\hat{c}_{21} -E^\alpha{}_\beta \frac{\d^2}{\d
b^\alpha_1 \d a_{2\beta}}\hat{c}_{12})\;,\quad i=1,2\,,
\\
\\

\ea \ee \be \label{Ho} \ba{c} \dps \hat{H}_{O}=
\dps\alpha_3(p,q,t) E_{\alpha\beta} \frac{\d^2}{\d a_{1\alpha} \d
a_{2\beta}}\hat{b}_{12}\hat{c}_{12} +\beta_3(p,q,t)
E^{\alpha\beta} \frac{\d^2}{\d b_1^\alpha \d
b_2^\beta}\hat{a}_{12}\hat{c}_{12}
\\
\\
\dps+\gamma_3(p,q,t) E_\alpha{}^\beta \frac{\d^2}{\d a_{1\alpha}
\d b_2^\beta}\;.
\\

\ea \ee Here $E^{\alpha\beta}=DV^{\alpha\beta}$ is the frame field
(see Appendix A). The coefficients $\alpha,\;\beta,\;\gamma$,
which parameterize various types of index contractions, depend on
the operators: \be \label{t} p=\hat{a}_{12}\hat{b}_{12}\;,\qquad
q=\hat{c}_{12}\hat{c}_{21}\;,\qquad t=\hat{c}_{11}\hat{c}_{22}\;,
\ee where \be \label{abg} \ba{ccc} \dps\hat{a}_{12} =
V_{\alpha\beta}\frac{\d^2}{\d a_{1\alpha} \d a_{2\beta}}\;,&
\qquad\dps \hat{b}_{12} = V^{\alpha\beta}\frac{\d^2}{\d b_1^\alpha
\d b_2^\beta}\;,& \qquad \dps\hat{c}_{ij} = \frac{\d^2}{\d
a_{i\alpha} \d b_j^\alpha}\;.
\\
\\
\ea \ee In what follows we will use the notation ${\cal
A}_{\alpha, \beta, \gamma}(F,G)$ for (\ref{act}) with the
collective coefficients \be \label{collepar} \ba{c}
\alpha =(\alpha_1 , \alpha_2, \alpha_3)\;, \\
\beta = (\beta_1, \beta_2, \beta_3)\;, \\
\gamma = (\gamma_1, \gamma_2, \gamma_3)\,\;.\\

\ea \ee

In our analysis the higher spin gauge fields will be allowed to
take values in some associative (e.g., matrix) algebra $\Omega
\rightarrow \Omega_I{}^J$. The resulting ambiguity is equivalent
to the ambiguity of a particular choice of the Yang-Mills gauge
algebra in the spin 1 sector. The classification of the higher
spin gauge theories associated with the different Yang-Mills
algebras is given in \cite{Vc}.  Therefore, the higher spin
actions (\ref{bosact2}) and (\ref{fermact}) are formulated in
terms of the trace $tr$ in this matrix algebra (to be not confused
with the trace in the star product algebra). As a result, only
cyclic permutations of the matrix factors will be allowed under
the trace operation. Remarkably, this property  simplifies
considerably the analysis of the gauge invariance of the cubic
action. Note that the gravitational field is required to take
values in the center of the matrix algebra, being proportional to
the unit matrix. For this reason, the factors associated with the
gravitational field are usually written outside the trace.

For general coefficients, the quadratic part of the action
(\ref{acta}) does not describe massless higher spin fields because
of ghost-type degrees of freedom associated with extra fields
$\omega^{s,\:t}\,,t>0$. To eliminate these extra degrees of
freedom one should fix the operators $\hat{H}$ (\ref{He}) and
(\ref{Ho}) in a specific way by requiring the variation of the
quadratic action with respect to the extra fields to vanish
identically \cite{LV,vf}. This condition is referred to as the
{\it extra field decoupling condition}. Another restriction on the
form of the action (\ref{acta}) comes from the requirement that
its quadratic part should decompose into an infinite sum of free
actions for different copies of fields of the same spin associated
with the spinor traces. This {\it factorization condition}
\cite{VD5} fixes a convenient basis in the space of fields rather
than imposes true dynamical limitations on form of the action.
Both of these conditions on the form of the quadratic part of the
action (\ref{act}) are analyzed in section \ref{Quadratic Action}.
Also, we introduce the {\it C-invariance condition} \cite{VD5}
which states that the action (\ref{act}) possesses the cyclic
property with respect to the central element of the higher spin
superalgebra. Being imposed, this natural
 condition simplifies greatly the analysis of
the dynamical system involving infinite sequences of
supermultiplets of the same spin. We show that the {\it
factorization condition} along with the {\it extra field
decoupling condition} and the {\it C-invariance condition} fix the
functions $\alpha,\beta,\gamma$ (\ref{collepar}) up to the
normalization coefficients in front of the individual free actions
modulo some ambiguity associated with total derivative terms in
the Lagrangian.

In the sequel we find a precise form of the cubic action
(\ref{acta}) that describes properly higher-spin-gravi\-tatio\-nal
interactions of spin  $s\geq 3/2$ fields in the first nontrivial
order. Note, that although this positive result indicates the
existence of a full nonlinear higher spin action, the constructed
cubic action is not expected to be complete even at the cubic
level. As mentioned in Introduction one reason for this is that
the full spectrum of fields in the appropriate higher spin
supermultiplet also contains spin 0, 1/2 and 1 massless fields not
included in the consideration of this paper. Our modest goal here
is to show that, similarly to the $4d$ case \cite{FV1}, the
problem with (cubic) higher-spin-gravitational interactions in the
flat background \cite{diff,diff2} can be avoided in $\ads$.

As explained in \cite{FV1,VD5} the analysis of the gauge
invariance in the cubic order is simplified greatly by using the
First On-Mass-Shell Theorem. The condition that the higher spin
action is invariant under some deformation of the higher spin
gauge transformations is {\it equivalent} to the
condition\footnote{Note that terms resulting from the gauge
transformations of the gravitational fields and the compensator
$V^{\alpha\beta}$ contribute into the factors in front of the
higher spin curvatures in the action
$(\ref{act})-(\ref{fermact})$. The proof of the respective
invariances is given in \cite{VD5} and is based entirely on the
explicit $su(2,2)$ covariance and invariance of the whole
framework under diffeomorphisms. Also, one has to take into
account that the higher spin gauge transformation of the
gravitational fields is at least linear in the dynamical fields
and therefore has to be discarded in the analysis of $\Omega^2
\epsilon $ type terms under consideration.} that the original
(i.e. undeformed) higher spin gauge variation of the action is
zero once the linearized higher spin curvatures $R_1$ are replaced
by the Weyl tensors $C$ according to
(\ref{onmassO10})-(\ref{onmassO21}). As a result, the problem is
to find such functions $\alpha, \beta$ and $\gamma$
(\ref{collepar}) that \be \label{SEtr} \delta {\cal S} (R,R)  \Big
|_{E=h, R = h\wedge h C} \equiv  {\cal A}^h_{\alpha,\beta,\gamma}
(R,[R,\epsilon ]_\star)
 \Big |_{R = h\wedge h C} =0\,
\ee for an arbitrary gauge parameter $\epsilon(a,b,\psi,\bpsi|x)$.
As shown in section \ref{unred} this condition, supplemented with
the {\it factorization condition} along with the {\it extra field
decoupling condition} and the {\it C-invariance condition}, fixes
the coefficients in the form \be \label{gab11} \alpha_1(p,q,t) +
\beta_1(p,q,t)  =  \Phi_0  \sum_{m,n =0}^\infty (-1)^{m+n}
\frac{m+1}{2^{2(m+n+1)} (m+n+2)! m! (n+1)!} p^n q^m\,, \ee

\be \label{gab12} \gamma_1(p,q,t)  = \gamma_1(p+q) \,,\quad
\gamma_1(p)=\Phi_0  \sum_{m =0}^\infty (-1)^{m+1}
\frac{1}{2^{2m+3} (m+2)! m!} p^m\,, \ee

\be \label{gab21} \alpha_2(p,q,t) + \beta_2(p,q,t)  =
\frac{1}{4}(\alpha_1(p,q,t) + \beta_1(p,q,t))\,, \quad
\gamma_2(p,q,t)  = \frac{1}{4}\gamma_1(p,q,t)\,, \ee

\be \label{gab31}
\alpha_3(p,q,t)+\beta_3(p,q,t)=\Phi_0\sum_{m,n=0}^{\infty}(-1)^{m+n+1}\frac{1}{2^{2(m+n)+3}\,(m+1)!\,(m+n+2)!\,n!}\,p^m\,q^n\;,
\ee

\be \label{gab32}
\gamma_3(p,q,t)=\gamma_3(p+q)\;,\quad\gamma_3(p)=\Phi_0\sum_{m=0}^{\infty}(-1)^{m+1}\frac{1}{2^{2m+1}\,m!\,(m+1)!}\,p^m\;,
\ee where $\Phi_0 $ is an arbitrary normalization factor to be
identified with the (appropriately normalized in terms of the
cosmological constant) gravitational coupling constant.

\subsection{Quadratic Action}
\label{Quadratic Action}

The free part ${\cal S}_2$ of the action is obtained by the
substitution of the linearized curvatures and the background frame
field into (\ref{acta}). The resulting action is manifestly
invariant under the linearized transformations (\ref{lintr})
because the linearized curvatures $R_1$ are invariant i.e.,
$\delta R_1=0$. We want the free action to be a sum of actions for
the irreducible higher spin fields. This requirement is not
completely trivial because of the infinite degeneracy of the
algebra due to the traces.

The {\it factorization condition} requires \be \label{fc} {\cal
S}_2 = \sum_{n,\:s=0}^{\infty} {\cal B}_2^{s,n}
(\Omega_{E_{1,2}}^{n,s+2}) + \sum_{n,\:s=0}^{\infty} {\cal
F}_2^{s+3/2,n} (\Omega_{O_{1,2}}^{n,s+3/2})\;, \ee i.e. the terms
containing products of the fields  $\Omega^{n,\,s}$ and
$\Omega^{m,\,s}$ with $n\neq m$ in the trace decomposition
(\ref{tr1})-(\ref{tr2}) should all vanish. As follows from
(\ref{tr1})-(\ref{7okt.}) this is true iff \be \label{fact0} {\cal
A}_{\alpha,\:\beta,\:\gamma}(F,(T^+)^k G) ={\cal
A}_{\alpha^{(k)},\: \beta^{(k)},\: \gamma^{(k)}}((T^-)^k F,G)\;,
\qquad \forall k \ee for some new parameters $\alpha^{(k)},\:
\beta^{(k)},\: \gamma^{(k)}$ (\ref{collepar}). The {\it
factorization condition} for the bosonic action ${\cal B}$
(\ref{bosact}) was analyzed in \cite{VD5}, where it was shown that
\be \label{fc_boson} \ba{l} {\cal
B}_{\;\alpha,\,\beta\,,\gamma}(F_{E},T^+G_{E})= {\cal
B}_{\;\alpha^{(1)},\beta^{(1)},\gamma^{(1)}}(T^-F_{E},G_{E})\;,
\ea \ee where the new parameters $\alpha^{(1)},\beta^{(1)},
\gamma^{(1)}$ express unambiguously in terms of
$\alpha,\beta,\gamma$: \be \label{AA3} \alpha_{1,2}^{(1)} = \
4\left((2+p\frac{\d}{\d p})\frac{\d}{\d p}+ (1+q\frac{\d}{\d
q})\frac{\d}{\d q}+(2p\frac{\d}{\d p}+2q\frac{\d}{\d q}+
t\frac{\d}{\d t}+6)\frac{\d}{\d t}\right) \alpha_{1,2}\,, \ee \be
\label{BB3} \beta_{1,2}^{(1)} = 4\left((2+p\frac{\d}{\d
p})\frac{\d}{\d p}+ (1+q\frac{\d}{\d q})\frac{\d}{\d q}+
(2p\frac{\d}{\d p} + 2q\frac{\d}{\d q}+ t\frac{\d}{\d
t}+6)\frac{\d}{\d t}\right)\beta_{1,2}\,, \ee \be \label{GG3}
\gamma_{1,2}^{(1)} =4\left((1+p\frac{\d}{\d p})\frac{\d}{\d p}+
(2+q\frac{\d}{\d q})\frac{\d}{\d q}+ (2p\frac{\d}{\d p} +
2q\frac{\d}{\d q}+ t\frac{\d}{\d t}+6)\frac{\d}{\d
t}\right)\gamma_{1,2}\, \ee provided that the following relation
is satisfied \be \label{fact_0} (1+p\frac{\d}{\d p})(\alpha_{1,2}
+ \beta_{1,2})+ 2(1+q\frac{\d}{\d q})\gamma_{1,2} =0 \;. \ee As
observed in \cite{VD5}, (\ref{fact_0}) is automatically  true for
the coefficients $\alpha_{1,2}^{(1)},\beta_{1,2}^{(1)} $ and $
\gamma_{1,2}^{(1)}$ and, therefore, (\ref{fact_0}) guarantees
(\ref{fact0}) in the bosonic sector for all $k$.

In the fermionic sector one gets \be \label{12nov} \ba{l} {\cal
F}_{\alpha_3,\,\beta_3,\,\gamma_3,\,}(F_{O},T^+ G_{O}) = {\cal
F}_{\alpha_3^{(1)},\,\beta^{(1)}_3,\,\gamma^{(1)}_3,}(T^-
F_{O},G_{O})
\\
\\
\dps
 + \frac{1}{2}\int_{{\cal M}^5}Q_{O}(p,q,t)E_\alpha{}^\beta \frac{\d^2}{\d
a_{1\alpha} \d b^\beta_1}\:\hat{c}_{12}\; \wedge  {\rm
tr}(G_{O_2}(a_1,b_1)\wedge F_{O_1}(a_2,b_2))|_{a_i=b_i=0}
\\
\\
\dps
 + \frac{1}{2}\int_{{\cal M}^5}Q_{O}(p,q,t)E_\alpha{}^\beta \frac{\d^2}{\d
a_{1\alpha} \d b^\beta_1}\:\hat{c}_{12}\; \wedge  {\rm
tr}(F_{O_2}(a_1,b_1)\wedge G_{O_1}(a_2,b_2))|_{a_i=b_i=0}\;,
\\
\\
\ea \ee where \be Q_{O} = (1+p\frac{\d}{\d p})(\alpha_3 +
\beta_3)+\frac{\d}{\d q}\gamma_3\; \ee and \be \label{A3}
\alpha_3^{(1)} = 4\left((2+p\frac{\d}{\d p})\frac{\d}{\d p}+
(2+q\frac{\d}{\d q})\frac{\d}{\d q}+(2p\frac{\d}{\d
p}+2q\frac{\d}{\d q}+ t\frac{\d}{\d t}+7)\frac{\d}{\d t}\right)
\alpha_3\,, \ee \be \label{B3} \beta_3^{(1)} =
4\left((2+p\frac{\d}{\d p})\frac{\d}{\d p}+ (2+q\frac{\d}{\d
q})\frac{\d}{\d q}+ (2p\frac{\d}{\d p} + 2q\frac{\d}{\d q}+
t\frac{\d}{\d t}+7)\frac{\d}{\d t}\right)\beta_3\,, \ee \be
\label{G3} \gamma_3^{(1)} =4\left((1+p\frac{\d}{\d p})\frac{\d}{\d
p}+ (1+q\frac{\d}{\d q})\frac{\d}{\d q}+ (2p\frac{\d}{\d p} +
2q\frac{\d}{\d q}+ t\frac{\d}{\d t}+5)\frac{\d}{\d
t}\right)\gamma_3\,, \ee The {\it factorization condition}
therefore requires \be \label{fact} Q_{O} = (1+p\frac{\d}{\d
p})(\alpha_3 + \beta_3)+\frac{\d}{\d q}\gamma_3 =0 \;. \ee

{}From (\ref{fact}) it follows that the same relation  is true for
the coefficients $\alpha^{(1)}$, $\beta^{(1)}$ and $\gamma^{(1)}$
(\ref{A3})-(\ref{G3}), and, therefore, (\ref{fact_0}),
(\ref{fact}) guarantee (\ref{fact0}) for all $k$.

An important role in the analysis of \cite{VD5} was played by the
{\it C-invariance condition} requiring that $ {\cal B} (T^+\star
F_E,G_E) = {\cal B}(F_E,G_E\star T^+).$ In the purely bosonic case
the operator $T^+$ coincides with the central element $N$. The
meaning of the {\it C-invariance condition} is that the bilinear
form used for the construction of the action has the cyclic
(trace) property with respect to elements of the center of the
algebra. It simplifies greatly the analysis of interactions and,
eventually, allows for elementary reduction to the quotient
algebra with the ideal generated by the central element $N$
factored out (see section \ref{redmod}). The supersymmetric {\it
C-invariance condition} has analogous form \be \label{fcic} {\cal
A} (N\star F,G) = {\cal A}(F,G\star N)\;, \ee where $F$ and $G$
are any elements satisfying $F\star N= N\star F$, $G\star N=
N\star G$.  Making use of the formula \be \label{razl} N\star F =
(P^+-P^-)F \ee \be \ba{c} \dps
=((T^+-T^-)F_{E_1}-\frac{1}{4}F_{E_2})+(T^+-T^-)F_{O_1}\psi
\\
\\
\dps +(T^+-T^-)F_{O_2}\bpsi+ ((T^+-T^-)F_{E_2}-F_{E_1})\psi\bpsi
\ea \ee and taking into account the {\it factorization condition}
(\ref{fact0}), we rewrite the {\it C-invariance condition}
(\ref{fcic}) as \be \ba{c} \dps {\cal
B}_{\;\alpha,\,\beta\,,\gamma}(F_{E}, T^-G_{E})+ {\cal
B}_{\;\alpha^{(1)},\beta^{(1)},\gamma^{(1)}}(F_{E}, T^-G_{E})
\\
\\
\dps +{\cal F}_{\;\alpha,\,\beta\,,\gamma}(F_{O}, T^-G_{O}) +{\cal
F}_{\;\alpha^{(1)},\beta^{(1)},\gamma^{(1)}}(F_{O}, T^-G_{O})
\\
\\
\dps={\cal B}_{\;\alpha,\,\beta\,,\gamma}(T^-F_{E}, G_{E})+ {\cal
B}_{\;\alpha^{(1)},\beta^{(1)},\gamma^{(1)}}(T^-F_{E}, G_{E})
\\
\\
\dps +{\cal F}_{\;\alpha,\,\beta\,,\gamma}(T^-F_{O}, G_{O}) +{\cal
F}_{\;\alpha^{(1)},\beta^{(1)},\gamma^{(1)}}(T^-F_{O}, G_{O})
\\
\\
\dps-\frac{1}{4}{\cal
B}_{\;\alpha,\,\beta\,,\gamma}^{\prime}(F_{E_1},G_{E_2}) +{\cal
B}_{\;\alpha,\,\beta\,,\gamma}^{\prime\prime}(F_{E_1},G_{E_2})
\\
\\
+\dps\frac{1}{4}{\cal
B}_{\;\alpha,\,\beta\,,\gamma}^{\prime}(F_{E_2},G_{E_1})- {\cal
B}_{\;\alpha,\,\beta\,,\gamma}^{\prime\prime}(F_{E_2},G_{E_1})\,.

\ea \ee The condition is true iff \be \label{abc1} {\cal
B}_{\;\alpha,\,\beta\,,\gamma}(F_{E},G_{E}) = -{\cal
B}_{\;\alpha^{(1)},\beta^{(1)},\gamma^{(1)}}(F_{E},G_{E})\,, \ee
\be \label{abc3} {\cal F}_{\;\alpha,\,\beta\,,\gamma}(F_{O},G_{O})
= -{\cal
F}_{\;\alpha^{(1)},\beta^{(1)},\gamma^{(1)}}(F_{O},G_{O})\,, \ee
\be \label{tsr} \frac{1}{4}{\cal
B}_{\;\alpha,\,\beta\,,\gamma}^{\prime}(F_{E},G_{E}) ={\cal
B}_{\;\alpha,\,\beta\,,\gamma}^{\prime\prime}(F_{E},G_{E})\;, \ee
i.e. \be \label{f1} \ba{cc} \alpha_i(p,q,t)
=-\alpha_i^{(1)}(p,q,t)\;,&\qquad \beta_i(p,q,t)
=-\beta_i^{(1)}(p,q,t)\;, \qquad i=1,2,3\,,
\\
\\
\gamma_j(p,q,t) =-\gamma_j^{(1)}(p,q,t)\;,&\qquad \gamma_j(p,q,t)
=-\gamma_j^{(1)}(p,q,t)\;,\qquad j=1,2,3\; \ea \ee and \be
\alpha_2(p,q,t)=\frac{1}{4}\alpha_1(p,q,t)\,,\qquad
\dps\beta_2(p,q,t)=\frac{1}{4}\beta_1(p,q,t)\,, \qquad \dps
\gamma_2(p,q,t)=\frac{1}{4}\gamma_1(p,q,t)\,. \ee The conditions
(\ref{abc1})-(\ref{abc3}) are equivalent to the requirement that
the operators $T^-$ and $T^+$ satisfy the following conjugation
rules \be \label{Bconj} {\cal B}(T^\pm F_{E}, G_{E})=-{\cal B}(
F_{E},T^\mp G_{E})\,, \ee \be \label{Fconj} {\cal F}(T^\pm F_{O},
G_{O})=-{\cal F}( F_{O},T^\mp G_{O})\,. \ee It is worth to note
that the relations (\ref{Bconj})-(\ref{Fconj}) may be equivalently
represented in the form \be \label{cinvbos} {\cal B}(T^+\star
G_E,F_E)={\cal B}( G_E,F_E \star T^+)\,, \ee \be \label{cinvfer}
{\cal F}(T^+\star  G_O,F_O)={\cal F}( G_O,F_O \star T^+)\,, \qquad
{\cal F}( G_O\star  T^+,F_O)={\cal F}( G_O,T^+\star F_O )\,, \ee
as one can easily see using that \be \label{again25} T^+ \star
F_{E_{1,2}} =  \left (T^+ -T^- \right )F_{E_{1,2}}\;,\qquad [T^+,
F_{E_{1,2}}]_\star =0\,, \ee \be \label{N*R} T^+ \star F_{O_{1,2}}
= \left (T^+ -T^-  +\frac{1}{2} S^0\right )F_{O_{1,2}}\;, \ee \be
\label{R*N} F_{O_{1,2}} \star  T^+ = \left (T^+ -T^-  -\frac{1}{2}
S^0\right ) F_{O_{1,2}}\,, \ee \be \label{comrelT} [T^+,
F_{O_1}]_\star =F_{O_1}\;,\qquad [T^+, F_{O_2}]_\star =-F_{O_2}\;.
\ee Using (\ref{again25})-(\ref{comrelT}) along with
(\ref{sl2inv}) it is elementary to compute the relative
coefficients for the different copies of fields in the
decomposition (\ref{tr1})-(\ref{tr2}). The normalization
coefficients (\ref{coe})-(\ref{coe1}) are chosen so that the
linearized actions have the same form for different copies of the
higher spin fields parameterized by the label $n$ \be
\label{S2summ} {\cal S}_2 = \sum_{n,\:s=0}^{\infty} {\cal B}_2^s
(\Omega_{E_{1,2}}^{n,s+2}) + \sum_{n,\:s=0}^{\infty} {\cal
F}_2^{s+3/2} (\Omega_{O_{1,2}}^{n,s+3/2})\;. \ee In  the
linearized approximation it is therefore enough to analyze the
situation for any fixed $n$. We confine ourselves to the case of
$\Omega^{s^\prime} = \Omega^{0,\,s^\prime}$, i.e. we will assume
in the rest of this section that $T^- \Omega^{s^\prime} = 0$.

The {\it extra field decoupling condition} requires \be
\label{exdc1} \frac{\delta {\cal B}_2}{\delta\omega_{e_{1,2}}^{t}}
\equiv 0\;, \quad {\rm for}\;\; t\geq 2\;, \quad {\rm and} \quad
\frac{\delta {\cal F}_2}{\delta\omega_{o_{1,2}}^{t}} \equiv 0\;,
\quad {\rm for}\;\; t\geq 1\;. \ee It was analysed in \cite{VD5}
for the bosonic sector and in \cite{A1} for free fermions. For the
reader's convenience we sketch here the main steps of this
analysis. The generic variation of ${\cal S}_2$ is schematically
\be \label{varenik} \ba{c} \dps \delta{\cal S}_2
=\frac{1}{2}\int_{{\cal M}^5} D_0\hat{H}_O\wedge
\delta\Omega_O\wedge R_{1,\;O} +\frac{1}{2}\int_{{\cal M}^5}
D_0\hat{H}_O\wedge
R_{1,\;O}\wedge \delta\Omega_O\\
\\\dps +\int_{{\cal M}^5} D_0\hat{H}_E\wedge
R_{1,\;E}\wedge \delta\Omega_E\;. \ea \ee According to
(\ref{expan1}) and (\ref{expan3}), (\ref{expan3per}) generic
variation of the extra fields has the form \be \label{extravarE}
\delta\Omega^{ex}_{E_{1,2}}(a,b)=(S^+)^2\xi_{E_{1,2}}(a,b)\,, \ee
with an arbitrary $\xi_{E_{1,2}}(a,b)$ satisfying
$(N_a-N_b-4)\xi_{E_{1,2}}(a,b)=0$, and \be \label{extravarO}
\delta\Omega^{ex}_{O_2}(a,b)=S^+\xi_{O_2}(a,b)\,,\qquad
\delta\Omega^{ex}_{O_1}(a,b)=S^-\xi_{O_1}(a,b)\,, \ee with
arbitrary $\xi_{O_{1,2}}(a,b)$ satisfying
$(N_b-N_a-3)\xi_{O_{1}}(a,b)=0$ and
$(N_a-N_b-3)\xi_{O_{2}}(a,b)=0$. The condition $\delta{\cal
S}_2=0$ with respect to the extra field variations
(\ref{extravarE}), (\ref{extravarO}) requires \be
\label{posleuzhina} \ba{l} \dps \alpha_i(p,q,0)+\beta_i(p,q,0) =
-2\int_0^1\: du \:(1+q\frac{\d}{\d q})\rho_i(pu+q,0)\;,\qquad
i=1,2\;,
\\
\\
\dps\alpha_3(p,q,0)+\beta_3(p,q,0) = -\int_0^1\: du\: \frac{\d}{\d
p}\rho_3(pu+q,0)\;,
\\
\\
\gamma_i(p,q,0)=\rho_i(p+q,0)\;,\qquad i=1,2,3\;.
\\
\\
\ea \ee Here the functions of one variable $\rho_i
(p+q)\,,i=1,2,3$ parameterize the leftover ambiguity in the
coefficients in front of the free actions of fields with different
spins.

As observed in \cite{VD5,A1}, at the free field level, there is an
ambiguity in the coefficients $\alpha_i(p,q,t)$ and
$\beta_i(p,q,t),\;i=1,2,3$ due to the freedom in adding a total
derivative \be \label{totder} \ba{c} \dps \delta {\cal S}_2 =
\frac{1}{2} \sum_{j=1}^{2} \int_{{\cal M}^5}d  \Big (  \Phi_j
(p,q,t ) {\rm tr} ( R_{E_j}(a_1 ,b_1 |x )\wedge R_{E_j} (a_2 ,b_2
|x ) )\Big|_{a_i=b_i=0}\Big )
\\
\\
+\dps \frac{1}{2}\int_{{\cal M}^5} d \Big(\Phi_3
(p,q)\hat{c}_{12}\, {\rm tr}( R_{O_2}(a_1,b_1)\wedge
R_{O_1}(a_2,b_2))\Big|_{a_i=b_i=0}\Big)
\\
\\
\\
\dps = \frac{1}{2} \sum_{j=1}^{2} \int_{{\cal M}^5} \frac{\d
\Phi_j (p,q,t)}{\partial p}\Big ( h^{\alpha\beta} \frac{\d^2}{\d
b_1^\alpha \d b_2^\beta}\hat{a}_{12} -
h_{\alpha\beta}\frac{\d^2}{\d a_{1\alpha}\d
a_{2\beta}}\hat{b}_{12}\Big)
\\
\\
\wedge {\rm tr}(R_{E_j} (a_1 ,b_1 |x )\wedge R_{E_j} (a_2 ,b_2 |x
)) \Big|_{a_i=b_i=0}
\\
\\
\dps + \frac{1}{2}\int_{{\cal M}^5} \frac{\d \Phi_3 (p,q)}{\d
p}\Big(h^{\alpha\beta} \frac{\d^2}{\d b_1^\alpha \d
b_2^\beta}\hat{a}_{12}\hat{c}_{12} - h_{\alpha\beta}
\frac{\d^2}{\d a_{1\alpha} \d a_{2\beta}}\hat{b}_{12}\hat{c}_{12}
\Big)
\\
\\
\dps \wedge {\rm tr}( R_{O_2}(a_1,b_1)\wedge
R_{O_1}(a_2,b_2))\Big|_{a_i=b_i=0}\,.
\\
\\

\ea \ee As a result, the variation of the coefficients \be
\label{nov2} \label{de} \delta \alpha_i (p,q,t) = \epsilon_i
(p,q,t)\,,\qquad \delta \beta_i (p,q,t) = -\epsilon_i
(p,q,t)\,,\quad i=1,2,3 \ee does not affect the physical content
of the quadratic action, i.e.,  in accordance with
(\ref{posleuzhina}), only the combination $\alpha_i
(p,q,t)+\beta_i (p,q,t)$ has invariant meaning at the free field
level.

Thus, the {\it factorization condition} (\ref{fact0}) along with
the {\it extra field decoupling condition} (\ref{exdc1}) fix the
functions $\alpha, \beta, \gamma$ (discarding the trivial
ambiguity (\ref{nov2})) up to arbitrary functions $\rho(p)$
parameterizing the ambiguity in the normalization coefficients in
front of the individual free bosonic and fermionic actions.
Remarkably, the analysis of the gauge invariance in the cubic
order fixes  the functions $\rho(p)$ unambiguously.

\subsection{Cubic Interactions}
\label{unred}

Now we are in a position to analyze the condition (\ref{SEtr}) to
prove the existence of a deformation of the higher spin gauge
transformation that leaves the cubic part of the action
(\ref{acta}) invariant up to higher-order corrections. The
undeformed higher spin transformation of the curvatures $\delta
R=[R,\epsilon]_\star$ with \be \label{r}
R(a,b,\psi,\bpsi|x)=R_{E_1}(a,b|x)+R_{O_1}(a,b|x)\psi+R_{O_2}(a,b|x)\bpsi+R_{E_2}(a,b|x)\psi\bpsi\;,
\ee \be
\epsilon(a,b,\psi,\bpsi|x)=\epsilon_{E_1}(a,b|x)+\epsilon_{O_1}(a,b|x)\psi+\epsilon_{O_2}(a,b|x)\bpsi+\epsilon_{E_2}(a,b|x)\psi\bpsi
\ee gives \be \label{re1} \delta R_{E_1} =
[R_{E_1},\epsilon_{E_1}]_\star+\frac{1}{4}[R_{E_2},\epsilon_{E_2}]_\star+
\frac{1}{2}[R_{O_1},\epsilon_{O_2}]_\star+\frac{1}{2}[R_{O_2},\epsilon_{O_1}]_\star\;\;,
\ee \be \label{ro1} \delta R_{O_1} =
[R_{E_1},\epsilon_{O_1}]_\star+[R_{O_1},\epsilon_{E_1}]_\star
-\frac{1}{2}\{R_{O_1},\epsilon_{E_2}\}_\star+\frac{1}{2}\{R_{E_2},\epsilon_{O_1}\}_\star\;\;,
\ee \be \label{ro2} \delta R_{O_2} =
[R_{E_1},\epsilon_{O_2}]_\star+[R_{O_2},\epsilon_{E_1}]_\star+
\frac{1}{2}\{R_{O_2},\epsilon_{E_2}\}_\star-\frac{1}{2}\{R_{E_2},\epsilon_{O_2}\}_\star\;\;,
\ee \be \label{re2} \delta R_{E_2} =
[R_{E_1},\epsilon_{E_2}]_\star+[R_{E_2},\epsilon_{E_1}]_\star+
\{R_{O_1},\epsilon_{O_2}\}_\star-\{R_{O_2},\epsilon_{O_1}\}_\star\;\;,
\ee where $[f,g]_\star=f\star g - g\star f$ and
$\{f,g\}_\star=f\star g + g\star f$ for $f=f(a,b)$ and $g=g(a,b)$.

As argued in \cite{VD5},  the gauge transformation deforms to \be
\label{skoro1} \delta\Omega =\delta^g\Omega+\Delta(R,\epsilon)\;,
\ee where $\Delta(R,\epsilon)$ denotes some $R$-dependent terms
such that $\Delta(0,\epsilon)=0$ and $\delta^g$ denotes the gauge
transformation (\ref{gotr}). The transformations (\ref{skoro1})
can be  rewritten as \be \label{skoro11} \ba{c} \delta\Omega_E
=(\delta^g\Omega)_E+\stackrel{\sim}{\Delta}_E(R_E,\epsilon_E)+
\stackrel{\approx}{\Delta}_E(R_O,\epsilon_O)\;,
\\
\\
\delta\Omega_O
=(\delta^g\Omega)_O+\stackrel{\sim}{\Delta}_O(R_O,\epsilon_E)+
\stackrel{\approx}{\Delta}_O(R_E,\epsilon_O)\;. \ea \ee {}Our aim
is to find an action ${\cal S}$ that admits a consistent
deformation of the gauge transformation guaranteeing that \be
\label{skoro2} \delta^g{\cal S}+\frac{\delta {\cal S}_2}{\delta
\omega^{ph}}\Delta\omega^{ph}
 = O(\Omega^3\epsilon)\;,
\ee where $\Delta$ is some deformation of transformation law of
the physical fields to be found. Taking into account
(\ref{skoro11}), the second term gets the form \be
\label{skoro_uzh} \ba{c} \dps \frac{\delta {\cal S}_2}{\delta
\omega^{ph}}\Delta\omega^{ph} = \dps \frac{\delta {\cal
B}_2}{\delta
\omega_e^{ph}}\stackrel{\sim}{\Delta}_E(R_E,\epsilon_E)
+\frac{\delta {\cal B}_2}{\delta
\omega_e^{ph}}\stackrel{\approx}{\Delta}_E(R_O,\epsilon_O)
\\
\\
+\dps \frac{\delta {\cal F}_2}{\delta
\omega_o^{ph}}\stackrel{\sim}{\Delta}_O(R_O,\epsilon_E)
+\frac{\delta {\cal F}_2}{\delta
\omega_o^{ph}}\stackrel{\approx}{\Delta}_O(R_E,\epsilon_O)\;.

\ea \ee Note that a deformation of the gauge variation of the
extra fields does not contribute into the variation to the order
under consideration because of (\ref{exdc1}). The first term on
the l.h.s. of (\ref{skoro2}) has the structure $R_E R_E \epsilon_E
+ R_E R_O \epsilon_O + R_O R_O \epsilon_E$, where all curvatures
are linearized. Imposing the constraints on the extra fields
proposed in \cite{LV,vf} which imply that the First-On-Mass-Shell
Theorem is satisfied we use the representation
(\ref{2nov})-(\ref{1nov}) for the linearized curvatures and
rewrite schematically the first term in (\ref{skoro2}) as \be
\label{uzhe} \ba{c} \dps C_E C_E \epsilon_E + C_E C_O \epsilon_O +
C_O C_O \epsilon_E
\\
\\
\dps + \stackrel{\sim}{{\cal H}}_E(R_E,\frac{\delta {\cal
B}_2}{\delta \omega_e^{ph}}, \epsilon_E) +
\stackrel{\approx}{{\cal H}}_E(R_O,\frac{\delta {\cal B}_2}{\delta
\omega_e^{ph}}, \epsilon_O)
\\
\\
+\dps   \stackrel{\sim}{{\cal H}}_O(R_O,\frac{\delta {\cal
F}_2}{\delta \omega_o^{ph}}, \epsilon_E) +
\stackrel{\approx}{{\cal H}}_O(R_E,\frac{\delta {\cal F}_2}{\delta
\omega_o^{ph}}, \epsilon_O)\;,

\ea \ee where ${\cal H}_E$ and ${\cal H}_O$ are some trilinear
functionals. Clearly, all terms in  ${\cal H}_E$ and ${\cal H}_O$
can be compensated by the appropriate deformations
$\stackrel{\sim}{\Delta}$ and $\stackrel{\approx}{\Delta}$. The
terms  bilinear in the higher spin Weyl tensors $C$ cannot be
compensated this way. The condition that the higher spin action is
invariant under some deformation of the higher spin
transformations is therefore equivalent to the requirement that
the $C^2$ terms cancel out. This is expressed by (\ref{SEtr}).

Let us start our analysis with the variation with respect to an
arbitrary bosonic higher spin transformation with the parameter
$\epsilon_{E_1}(a,b|x)
=\epsilon^{\alpha(s)}_{\beta(s)}(x)a_{\alpha(s)}b^{\beta(s)}$.
According to (\ref{SEtr}), our aim is to prove that there exist
such coefficient functions $\alpha,\beta$ and $\gamma$
(\ref{collepar}) satisfying the {\it {} C-invariance condition},
{\it factorization condition} and {\it extra field decoupling
condition} that \be \ba{l} \label{sinvmn0} \dps2{\cal B}^h \left
(R_{1,\;E_{1,2}}  , [\epsilon_{E_{1}} , R_{1,\;E_{1,2}} \,]_\star
\right )\Big |_{R=h\wedge h C}
\\
\\
\dps+ {\cal F}^h \left (R_{1,\;O_2} , [\epsilon_{E_1} ,
R_{1,\;O_1} \,]_\star \right )\Big |_{R=h\wedge h C}
\\
\\
\dps+{\cal F}^h \left ([\epsilon_{E_1} ,  R_{1,\;O_2} \,]_\star,
 R_{1,\;O_1} \right )\Big |_{R=h\wedge h C} =0
\ea \ee for arbitrary gauge parameter $\epsilon_{E_1} =
\epsilon_{E_1}(a,b|x)$ and arbitrary Weyl tensors $C(a)$. Taking
into account the decompositions (\ref{tr1_R})-(\ref{tr2_R}), the
condition (\ref{sinvmn0}) takes the form \be \ba{l} \label{sinvmn}
\dps2\sum_{mn} {\cal B}^h \left ((T^+)^m v_{E_{1,2},m} (T^0 )
R_{1,\;E_{1,2}}^m (a ,b ) , [\epsilon_{E_{1}} , (T^+)^n
v_{E_{1,2},n} (T^0 ) R_{1,\;E_{1,2}}^n (a ,b ) \,]_\star \right
)\Big |_{R=h\wedge h C}
\\
\\
\dps+\sum_{mn} {\cal F}^h \left ((T^+)^m v_{O_2,m} (T^0 )
R_{1,\;O_2}^m (a ,b ) , [\epsilon_{E_1} , (T^+)^n v_{O_1,n} (T^0 )
R_{1,\;O_1}^n (a ,b ) \,]_\star \right )\Big |_{R=h\wedge h C}
\\
\\
\dps+\sum_{mn} {\cal F}^h \left ([\epsilon_{E_1} , (T^+)^m
v_{O_2,m} (T^0 )  R_{1,\;O_2}^m (a ,b ) \,]_\star, (T^+)^n
v_{O_1,n} (T^0 ) R_{1,\;O_1}^n (a ,b )

\right )\Big |_{R=h\wedge h C} =0 \ea \ee with an arbitrary gauge
parameter $\epsilon_{E_1}(a,b|x)$ and arbitrary Weyl tensors
$C_{E_{1,2}}^n(a|x)$ and $C_{O_{1,2}}^m(a|x)$ in the decomposition
(\ref{onmassO10})-(\ref{onmassO21}) for the linearized higher spin
curvatures $R_{1,\,E_{1,2}}^n(a,b|x)$ and
$R_{1,\,O_{1,2}}^m(a,b|x)$.

First of all, one  observes that the dependence of $v_n (T^0 )$ on
$T^0$ can be absorbed into (spin-dependent) rescalings of the Weyl
tensors $C^n(a)$ which are treated as arbitrary field variables in
this consideration. As a result it is enough to prove
(\ref{sinvmn}) for arbitrary constant coefficients $v_n$. Now let
us show that, once (\ref{sinvmn})  is valid for  $m=n=0$, it is
automatically true for all other values of $m$ and $n$ as a
consequence of the relations (\ref{cinvbos}) and (\ref{cinvfer})
which follow from the {\it factorization condition} and the {\it
C-invariance condition}. For the bosonic part this was shown in
\cite{VD5}, where the proof was based on the bosonic {\it
C-invariance condition} (\ref{cinvbos}). Because the relation
(\ref{cinvbos}) is still valid, the proof remains the same.
Thereby we focus on the fermionic part ${\cal F}$.

Suppose that (\ref{sinvmn}) is true for $m_0 \geq m \geq 0$, $n_0
\geq n \geq 0$. Consider the term with $m= m_0 +1$. Then, from
(\ref{N*R}) it follows \be \label{inet!} \ba{c} \dps (T^+)^{m_0
+1}\, R^{m_{0+1}}_{1,\;O_2}(a ,b )= T^+ \star ((T^+)^{m_0}\,
R^{m_{0+1}}_{1,\;O_2}(a ,b ))
\\
\dps + T^-\,(T^+)^{m_0}\, R^{m_{0+1}}_{1,\;O_2}(a ,b
)-\frac{1}{2}(T^+)^{m_0}\, R^{m_{0+1}}_{1,\;O_2}(a ,b )\,.
\\
\ea \ee After the substitution of this expression into
(\ref{sinvmn}) the term containing $T^-  $ gives zero contribution
by the induction assumption since, taking into account that \\
$T^-\,R^{m_{0+1}}_{1,\;O_2}(a ,b)=0$, $T^-$ decreases a number of
$T^+$. The last term in (\ref{inet!}) does not contribute by the
induction assumption as well. By virtue of the (\ref{cinvfer})
along with the commutation relations (\ref{comrelT}) and the fact
that $T^+$ commutes with bosonic elements of $cu(1,1|8)$ the terms
containing the star product with $T^+$ are \be \ba{c} \dps {\cal
F}^h\left ((T^+)^{m_0}v_{O_2,m_0}\, (T^0)\,R^{m_0 }_{1,\;O_2}\, (a
,b )\Big |_{m.s.} , [T^+ \star  \epsilon_{E1} , (T^+)^{n_0}\,
v_{O_1,n_0} (T^0 )\, R^{n_0}_{1,\;O_1}\, (a ,b )\Big |_{m.s.}
\,]_\star \right )
\\
\\
+\dps {\cal F}^h\left ([T^+ \star  \epsilon_{E1} , (T^+)^{m_0}\,
v_{O_2,m_0} (T^0 )\,  R^{m_0}_{1,\;O_2}\, (a ,b )\Big |_{m.s.}
\,]_\star , (T^+)^{n_0}v_{O_1,n_0}\, (T^0)\,R^{n_0 }_{1,\;O_1}\,
(a ,b )\Big |_{m.s.} \right )
\\
\ea \ee which is zero by the induction assumption valid for any
$\epsilon_{E1}$. Analogously, one performs induction $n_0 \to n_0
+1$ with respect to $R_{1,\;O_1}$ with the help of
(\ref{cinvfer})-(\ref{comrelT}).

Thus it is sufficient to find the coefficients satisfying the {\it
C-invariance condition} and the {\it factorization condition } for
traceless curvatures $R={\cal R}\equiv R^0$ satisfying $T^-({\cal
R})=0$. In other words one has to prove that \be \label{eshe}
{\cal S}^h({\cal R}, [\epsilon_{E1}, {\cal R}]_\star)=0\,, \ee
where \be \label{onmassE} {\cal
R}_{E_{1,2}}(a,b)=H_{2\,\alpha}{}^\beta \frac{\d^2}{\d a_\alpha \d
b^\beta}{\rm Res}_\mu (C_{E_{1,2}}(\mu a+\mu^{-1}b))\, \ee for
bosons and \be \label{onmassO1} {\cal
R}_{O_1}(a,b)=H_{2\,\alpha}{}^\beta \frac{\d^2}{\d a_\alpha \d
b^\beta} {\rm Res}_\mu (\mu \,C_{O_1}(\mu a+\mu^{-1}b))\,, \ee \be
\label{onmassO2} {\cal R}_{O_2}(a,b)=H_{2\,\alpha}{}^\beta
\frac{\d^2}{\d a_\alpha \d b^\beta}{\rm Res}_\mu (\mu^{-1}
\,C_{O_2}(\mu a+\mu^{-1}b))\, \ee for fermions. Note that because
$T^-({\cal R})=0$, the terms containing $\hat{c}_{11}$ (\ref{abg})
and, therefore, $t$ (\ref{t}) do not contribute into the condition
(\ref{eshe}).

Consider the variation (\ref{sinvmn}) of the fermionic action: \be
\label{13nov} \ba{c} \dps \delta {\cal F}^h=\int_{{\cal M}^5}
\hat{H}_{O}\wedge  {\rm tr} (\delta \,{\cal
R}_{O_2}(a_1,b_1)\wedge {\cal R}_{O_1}(a_2,b_2))|_{a_i=b_i=0}
\\
\\
\dps+ \int_{{\cal M}^5} \hat{H}_{O}\wedge {\rm tr}( {\cal
R}_{O_2}(a_1,b_1)\wedge \delta \,{\cal
R}_{O_1}(a_2,b_2))|_{a_i=b_i=0}\,.
\\
\\

\ea \ee Substituting $\delta {\cal R}_{O_1} = [{\cal
R}_{O_1},\epsilon_{E_1}]_\star$ and $\delta {\cal R}_{O_2} =
[{\cal R}_{O_2},\epsilon_{E_1}]_\star $ one gets \be
\label{varrff} \ba{c} \dps\delta {\cal F}^h =\int_{{\cal M}^5}
\hat{H}_{O}\wedge  {\rm tr}(({\cal R}_{O_2}\star
\epsilon_{E_1})(a_1,b_1)\wedge {\cal
R}_{O_1}(a_2,b_2))|_{a_i=b_i=0}
\\
\\
\dps -\int_{{\cal M}^5} \hat{H}_{O}\wedge {\rm tr}(
(\epsilon_{E_1}\star {\cal R}_{O_2})(a_1,b_1)\wedge {\cal
R}_{O_1}(a_2,b_2))|_{a_i=b_i=0}
\\
\\
\dps+ \int_{{\cal M}^5} \hat{H}_{O}\wedge  {\rm tr}({\cal
R}_{O_2}(a_1,b_1)\wedge ({\cal R}_{O_1}\star
\epsilon_{E_1})(a_2,b_2))|_{a_i=b_i=0}
\\
\\
\dps - \int_{{\cal M}^5} \hat{H}_{O}\wedge  {\rm tr} ({\cal
R}_{O_2}(a_1,b_1)\wedge (\epsilon_{E_1}\star {\cal
R}_{O_1})(a_2,b_2))|_{a_i=b_i=0}\,.
\\
\\
\ea \ee Let us calculate explicitly the first term in
(\ref{varrff}). Making use of the star product (\ref{spa}) along
with the identities (\ref{id1})-(\ref{id2}) applied to the
background fields, and rewriting (\ref{onmassO1}),
(\ref{onmassO2}) as \be \label{onmassO1e} \left.\dps {\cal
R}_{O_1}(a,b)={\rm Res}_\mu\, \mu^{-1}e^{\mu^{-1}\,
a_\gamma\frac{\d}{\d c_\gamma}+\mu\, b^\gamma\frac{\d}{\d
c^\gamma}} H_2^{\alpha\beta}\frac{\d^2}{\d c^{\,\alpha} \d
c^{\,\beta}}C_{O_1}(c)\right|_{c=0}\,, \ee \be \label{onmassO2e}
\left.\dps {\cal R}_{O_2}(a,b)={\rm Res}_\mu\, \mu^{-1}e^{\mu\,
a_\gamma\frac{\d}{\d c_\gamma}+\mu^{-1}b^\gamma\frac{\d}{\d
c^\gamma}} H_2^{\alpha\beta}\frac{\d^2}{\d c^{\,\alpha} \d
c^{\,\beta}}C_{O_2}(c)\right|_{c=0}\,, \ee one finds \be
\label{varrf1} \ba{c} \dps \int_{{\cal M}^5} \hat{H}_{O}\wedge
{\rm tr}( ({\cal R}_{O_2}\star \epsilon_{E_1})(a_1,b_1)\wedge
{\cal R}_{O_1}(a_2,b_2))|_{a_i=b_i=0}
\\
\\
\dps =-\frac{1}{30} \int_{{\cal M}^5}\, H_5\,\bar{k}^2\,{\rm
Res}_\mu
\left(\mu^{-1}e^{\frac{1}{2}(\mu^{-1}\bar{u}_2-\mu\,\bar{v}_2)}(\mu\,\bar{k}-\bar{u}_1)\,\Phi(Y)\right)\times
\\
\\
\dps \times {\rm tr}
\left.(C_{O_2}(c_2)C_{O_1}(c_1)\epsilon_{E_1}(a_3,b_3))\right|_{a=b=c=0}\,,
\\
\\

\ea \ee where $H_{5}$ denotes the vacuum  $5$-form defined in
Appendix A, \be \bar{k}=\frac{\d^2}{\d c_{1\,\alpha} \d
c_2^\alpha}\,,\qquad \bar{u}_i=\frac{\d^2}{\d c_i^\alpha \d
a_{3\,\alpha}}\,,\qquad \bar{v}_i=\frac{\d^2}{\d c_{i\,\alpha} \d
b_3^\alpha}\,, \ee

\be \label{def_Y}
Y=(\mu^{-1}\bar{k}+\bar{v}_1)(\mu\,\bar{k}-\bar{u}_1) \ee and \be
\Phi(Y)=Y(\alpha_3(Y,-Y)+\beta_3(Y,-Y))+\gamma_3(Y,-Y)\,. \ee
Calculating analogously the remaining terms in (\ref{varrff}) one
obtains for the whole variation (\ref{13nov}) \be \ba{c} \dps
\delta {\cal F}^h=-\frac{1}{15} \int_{{\cal M}^5}\,
H_5\,\bar{k}^2\,{\rm Res}_\mu
\left(\mu^{-1}e^{\frac{1}{2}(\mu^{-1}\bar{u}_2-\mu\,\bar{v}_2)}
(\mu\,\bar{k}-\bar{u}_1)\,\Phi(Y)\right)\times
\\
\\
\times {\rm tr}
\left.(C_{O_2}(c_2)C_{O_1}(c_1)\epsilon_{E_1}(a_3,b_3))\right|_{a=b=c=0}
\\
\\
\dps +\frac{1}{15} \int_{{\cal M}^5}\, H_5\,\bar{k}^2\,{\rm
Res}_\mu
\left(\mu^{-1}e^{\frac{1}{2}(\mu^{-1}\bar{v}_1-\mu\,\bar{u}_1)}
(\mu\,\bar{k}-\bar{v}_2)\,\Phi(Z)\right) \times
\\
\\
\times {\rm tr}\left.
(C_{O_2}(c_2)C_{O_1}(c_1)\epsilon_{E_1}(a_3,b_3))\right|_{a=b=c=0}\,,
\\
\\

\ea \ee where \be \label{def_W}
Z=(\mu\bar{k}-\bar{v}_2)(\mu^{-1}\,\bar{k}+\bar{u}_2)\,. \ee
Introducing notations \be \label{splitABFD} \ba{cc}
A=(\mu\bar{k}-\bar{u}_1)\,,&
B=(\mu^{-1}\,\bar{k}+\bar{v}_1)\,,\\
& \\
F=(\mu\bar{k}-\bar{v}_2)\,,& D=(\mu^{-1}\bar{k}+\bar{u}_2)\,,\\
\ea \ee the problem amounts to the search for a such function
$\Phi(Y)$ that \be \label{4.83} \ba{c} \dps \bar{k}^2\,{\rm
Res}_\mu\,\left(\mu^{-1}\,A\,e^{\frac{1}{2}(\mu^{-1}\bar{u}_2-\mu\,\bar{v}_2)}\,\Phi(AB)-
\mu^{-1}\,F\,e^{\frac{1}{2}(\mu^{-1}\bar{v}_1-\mu\,\bar{u}_1)}\,\Phi(FD)\right)\times
\\
\\
\left.\dps {\rm
tr}(C_{O_2}(c_2)C_{O_1}(c_1)\epsilon_{E_1}(a_3,b_3))\right|_{a=b=c=0}
=0\,.
\\
\\
\ea \ee Defining $\tilde{\Phi}(A,B)=A\Phi(AB)$ one rewrites
(\ref{4.83}) as \be \label{ustal} \ba{c} \bar{k}^2\,{\rm
Res}_\mu\,\left(\mu^{-1}\,\,e^{\frac{1}{2}(\mu^{-1}\bar{u}_2-\mu\,\bar{v}_2)}\,\tilde{\Phi}(A,B)
-\mu^{-1}\,e^{\frac{1}{2}(\mu^{-1}\bar{v}_1-\mu\,\bar{u}_1)}\,\tilde{\Phi}(F,D)\right)\times
\\
\\
\left.{\rm tr}
(C_{O_2}(c_2)C_{O_1}(c_1)\epsilon_{E_1}(a_3,b_3))\right|_{a=b=c=0}
=0\,.
\\
\ea \ee Now one observes that the function
$\tilde{\Phi}(A,B)=\Phi_0{\rm
Res}_\nu(\nu^{-1}e^{\frac{1}{2}(\nu\,A+\nu^{-1}B)})$, where
$\Phi_0$ is some normalization constant, solves (\ref{ustal}).

As a result, the condition (\ref{eshe}) amounts to \be \ba{c} \dps
A(\alpha_3(A,-A)+\beta_3(A,-A))+\gamma_3(A,-A)= \Phi_0
\,A^{-1}{\rm Res}_\nu(\nu^{-1}e^{\frac{1}{2}(\nu\,A+\nu^{-1})})
\\
\\
\dps =\frac{\Phi_0}{2}\int_0^1 du \,{\rm Res}_\nu\,
e^{\frac{1}{2}(\nu^{-1}+\nu u A)}\,. \ea \ee Taking into
account(\ref{posleuzhina}) this is solved by \be \label{gamma3fin}
\gamma_3(p)=\frac{\Phi_0}{2}\int_0^1 du \,{\rm Res}_\nu\,
e^{\frac{1}{2}(-\nu^{-1}+\nu p u)}\, \ee and \be \label{alpha3fin}
(\alpha_3+\beta_3)(p,q)=\frac{\gamma_3(p+q)}{q}-
\frac{\Phi_0}{2q}\int_0^1 du \,{\rm Res}_\nu\,
e^{\frac{1}{2}(-\nu^{-1}+\nu (p u+q))}\,. \ee With the aid of
these expressions one can see that the following identities are
true \be \label{to1} \Big (p\frac{\d^2 }{\d p^2} +2\frac{\d}{\d p}
+\frac{1}{4} \Big ) \gamma_3 (p) = 0 \,, \ee \be \label{to2} \Big
( \Big ( 2 + p \frac{\d}{\d p} \Big ) \frac{\d}{\d p} + \Big ( 2 +
q \frac{\d}{\d q} \Big ) \frac{\d}{\d q} + \frac{1}{4} \Big )
(\alpha_3 (p,q,0)+ \beta_3 (p,q,0)) =0\,. \ee {}From these
identities and relations (\ref{A3})-(\ref{G3}) it follows then
that the {\it $C-$invariance condition} (\ref{fcic}) is satisfied
with \be \alpha_3 (p,q,t) + \beta_3 (p,q,t)= \alpha_3
(p,q,0)+\beta_3 (p,q,0)\,,\qquad \gamma_3 (p,q,t) = \gamma_3
(p,q,0)\,. \ee The power series expansion of the expressions
(\ref{gamma3fin})-(\ref{alpha3fin}) for $\gamma_3(p)$ and
$\alpha_3(p,q,0) +\beta_3(p,q,0)$ gives (\ref{gab31}) and
(\ref{gab32}).

Thus it is shown that the coefficient functions (\ref{gamma3fin})
and (\ref{alpha3fin}) satisfy the {\it factorization condition},
{\it $C-$invariance condition}, {\it extra field decoupling
condition} and the condition (\ref{SEtr}) in the fermionic sector.
The leftover ambiguity in the coefficients $\alpha_3(p,q,t)
+\beta_3 (p,q,t)$ and $\gamma_3 (p,q,t)$ reduces to the overall
factor $\Phi_0$ in front of the fermionic action ${\cal F}$.

The explicit form of the  coefficients of the bosonic action was
fixed in \cite{VD5} by the requirement of its invariance under the
(appropriately deformed) higher spin transformations with the
parameters
$\epsilon(a,b|x)=\epsilon^{\alpha(s)}_{\beta(s)}(x)a_{\alpha(s)}b^{\beta(s)}$.
The results of \cite{VD5} remain true in our model. The respective
coefficient functions  are \be \label{intgga} \gamma_i (p) =
\frac{\Phi_i}{4}\int^1_0 dv v \, {\rm Res}_\nu \Big ( \nu e^{\half
(-\nu^{-1} +\nu vp )} \Big )\,,\quad i=1,2\, \ee and \be
\label{intga} \alpha_i (p,q,0)+\beta_i (p,q,0) = 2 \gamma_i (p+q)
- \half\Phi_i \int^1_0 du \, {\rm Res}_\nu \Big ( \nu e^{\half
(-\nu^{-1} +\nu(u p +q)) }\Big)\,, \quad i=1,2\,, \ee where
$\Phi_1$ and $\Phi_2$ are arbitrary real constants.

The variation with respect to bosonic parameters
\\ $\epsilon_{E_2}(a,b,\psi,\bpsi|x)=\epsilon^{\alpha(s)}_{\beta(s)}(x)a_{\alpha(s)}b^{\beta(s)}\psi\bpsi$
relates the  coefficients $\Phi_1,\,\Phi_2$ as \be \label{1/4}
\Phi_2=\frac{1}{4}\Phi_1\; \ee and gives equations on the
fermionic coefficients equivalent to those that follow from the
variation with respect to $\epsilon_{E_1}(a,b|x)$ (\ref{varrff}).
Note that the condition (\ref{1/4}) derived by virtue of the gauge
symmetry gives the same relation  between bosonic coefficients
(\ref{tsr}) as fixed by the {\it C-invariance condition}
(\ref{fcic}).

Consider now the variation of the full  action with respect to
fermionic transformation with an arbitrary gauge parameter
$\epsilon_{O}(a,b,\psi,\bpsi|x)=\epsilon_{O_1}(a,b|x)\psi
+\epsilon_{O_2}(a,b|x)\bpsi$. Taking into account
(\ref{re1})-(\ref{re2}), one obtains \be \label{varrf} \ba{c} \dps
\delta{\cal A}^h=\int_{{\cal M}^5} \hat{H}_{E_1}\wedge  {\rm tr}(
{\cal R}_{E_1}(a_1,b_1)\wedge ({\cal R}_{O_1}\star\epsilon_{O_2}-
\epsilon_{O_2}\star {\cal R}_{O_1})(a_2,b_2))|_{a_i=b_i=0}
\\
\\
\dps +2\int_{{\cal M}^5} \hat{H}_{E2}\wedge {\rm tr}( {\cal
R}_{E2}(a_1,b_1)\wedge ({\cal R}_{O_1}\star
\epsilon_{O_2}+\epsilon_{O_2}\star {\cal
R}_{O_1})(a_2,b_2))|_{a_i=b_i=0}
\\
\\
\dps+ \int_{{\cal M}^5} \hat{H}_{O}\wedge {\rm tr}( ({\cal
R}_{E_1}\star \epsilon_{O_2} -\epsilon_{O_2}\star {\cal
R}_{E_1})(a_1,b_1)\wedge {\cal R}_{O_1}(a_2,b_2))|_{a_i=b_i=0}
\\
\\
\dps -\frac{1}{2} \int_{{\cal M}^5} \hat{H}_{O}\wedge {\rm tr}(
({\cal R}_{E2}\star \epsilon_{O_2}+\epsilon_{O_2}\star {\cal
R}_{E2}) (a_1,b_1))\wedge {\cal R}_{O_1}(a_2,b_2))|_{a_i=b_i=0}
\\
\\
{\rm +\;\; analogous\;\; terms\;\; containing
\;\;\epsilon_{O_1}}\,.
\\
\\
\ea \ee Proceeding analogously to the bosonic transformation one
arrives at: \be \label{predvar} \ba{c} \dps \delta{\cal
A}^h=\frac{1}{15}\dps \int_{{\cal M}^5}\,H_5 \,\bar{k}^2\, {\rm
Res}_\mu\,e^{\frac{1}{2}(\mu^{-1} \bar{u}_2 -\mu\bar{v}_2)}\,
\left(\frac{1}{2}\Psi_1(Y)\frac{\d}{\d c_2^\sigma} -\mu(\mu^{-1}
\bar{k}+\bar{v}_1)\frac{\d \Psi_1(Y)}{\d Y}\frac{\d}{\d
c_1^\sigma}\right)
\\
\\
\times {\rm
tr}(\epsilon_{O_2}{}^{\sigma\gamma(s)}_{\;\rho(s)}a_{3\;\gamma(s)}b_3^{\rho(s)}
\left.\,C_{E_1}(c_1)C_{O_1}(c_2))\right|_{\,a=b=c=0}
\\
\\
\dps+\frac{1}{15}\int_{{\cal M}^5}\,H_5 \,\bar{k}^2\, {\rm
Res}_\mu\,e^{\frac{1}{2}(\mu^{-1} \bar{v}_1 -\mu\bar{u}_1)}\,
\left((Z\frac{\d \Phi(Z)}{\d Z}+\Phi(Z))\frac{\d}{\d c_2^\sigma}
-\frac{\mu}{2}(\mu^{-1} \bar{k}+\bar{u}_2)\Phi(Z)\frac{\d}{\d
c_1^\sigma}\right)
\\
\\
\times {\rm
tr}(\epsilon_{O_2}{}^{\sigma\gamma(s)}_{\;\rho(s)}a_{3\;\gamma(s)}b_3^{\rho(s)}
\left.\,C_{E_1}(c_1)C_{O_1}(c_2))\right|_{\,a=b=c=0}\;
\\
\\
+\dps \frac{2}{15}\int_{{\cal M}^5}\,H_5 \,\bar{k}^2\, {\rm
Res}_\mu\,e^{\frac{1}{2}(\mu^{-1} \bar{u}_2 -\mu\bar{v}_2)}\,
\left(\frac{1}{2}\Psi_2(Y)\frac{\d}{\d c_2^\sigma} -\mu(\mu^{-1}
\bar{k}+\bar{v}_1)\frac{\d \Psi_2(Y)}{\d Y}\frac{\d}{\d
c_1^\sigma}\right)
\\
\\
\times {\rm
tr}(\epsilon_{O_2}{}^{\sigma\gamma(s)}_{\;\rho(s)}a_{3\;\gamma(s)}b_3^{\rho(s)}
\left.\,C_{E_2}(c_1)C_{O_1}(c_2))\right|_{\,a=b=c=0}
\\
\\
\dps+\frac{1}{30}\int_{{\cal M}^5}\,H_5 \,\bar{k}^2\, {\rm
Res}_\mu\,e^{\frac{1}{2}(\mu^{-1} \bar{v}_1 -\mu\bar{u}_1)}\,
\left((Z\frac{\d \Phi(Z)}{\d Z}+\Phi(Z))\frac{\d}{\d c_2^\sigma}
-\frac{\mu}{2}(\mu^{-1} \bar{k}+\bar{u}_2)\Phi(Z)\frac{\d}{\d
c_1^\sigma}\right)
\\
\\
\times {\rm
tr}(\epsilon_{O_2}{}^{\sigma\gamma(s)}_{\;\rho(s)}a_{3\;\gamma(s)}b_3^{\rho(s)}
\left.\,C_{E_2}(c_1)C_{O_1}(c_2))\right|_{\,a=b=c=0}
\\
\\
\\
{\rm +\;\; analogous\;\; terms\;\; containing
\;\;\epsilon_{O_1}}\,,
\\
\\
\ea \ee where \be \label{ferbos} \ba{l} \dps \Psi_i(Y)=\Phi_i{\rm
Res}_\nu e^{\frac{1}{2}(\nu^{-1}+\nu Y)}\,,\quad i=1,2\,,
\\
\\
\dps \Phi(Z)=\Phi_0 Z^{-1}{\rm Res}_\nu\
(\nu^{-1}e^{\frac{1}{2}(\nu^{-1}+\nu Z)}) \ea \ee and $Y$ and $Z$
are defined by (\ref{def_Y}) and (\ref{def_W}). An important
observation is that the functions (\ref{ferbos}) satisfy \be
\label{connect} Z\frac{\d \Phi(Z)}{\d
Z}+\Phi(Z)=\frac{\Phi_0}{2\Phi_i}\,\Psi_i(Z)\,, \quad i=1,2\,. \ee
Using notations (\ref{splitABFD}) we get from (\ref{ferbos}) \be
\label{express} \ba{c} \dps D\Phi(DF)=\Phi_0{\rm
Res}_{\nu}(\nu^{-1}e^{\frac{1}{2}(\nu D +\nu^{-1}F)})\,,
\\
\\
\dps B\,\d\Psi_i(AB)=\frac{\Phi_i}{2}{\rm
Res}_{\nu}(\nu^{-1}e^{\frac{1}{2}(\nu B +\nu^{-1}A)})\,,\quad
i=1,2\,. \ea \ee Assuming the relation between bosonic
coefficients (\ref{1/4}), with the help of (\ref{connect})  the
problem is reduced to the search of a solution to the equations
\be \label{4.100} \ba{c} \dps \frac{1}{2}\bar{k}^2\,{\rm
Res}_{\mu,\,\nu}\,(e^{\frac{1}{2}(\mu^{-1} \bar{u}_2
-\mu\bar{v}_2)}\,\Psi_1(Y)+
\frac{\Phi_0}{\Phi_1}\,e^{\frac{1}{2}(\mu^{-1} \bar{v}_1
-\mu\bar{u}_1)}\,\Psi_1(Z))\frac{\d}{\d c_2^\sigma}\times
\\
\\
\times {\rm
tr}(\epsilon_{O_2}{}^{\sigma\gamma(s)}_{\;\rho(s)}a_{3\;\gamma(s)}b_3^{\rho(s)}
\left.\,C_{E_{1,2}}(c_1)C_{O_1}(c_2))\right|_{\,a=b=c=0}
\\
\\
- \dps \bar{k}^2\,{\rm Res}_{\mu,\,\nu}\,
(\mu\,e^{\frac{1}{2}(\mu^{-1} \bar{u}_2 -\mu\bar{v}_2)}\,B\,\d
\Psi_1(AB)+ \frac{1}{2}\, \mu\,\,e^{\frac{1}{2}(\mu^{-1} \bar{v}_1
-\mu\bar{u}_1)}\,D\,\Phi(FD))\frac{\d}{\d c_1^\sigma}\times
\\
\\
\times {\rm
tr}(\epsilon_{O_2}{}^{\sigma\gamma(s)}_{\;\rho(s)}a_{3\;\gamma(s)}b_3^{\rho(s)}
\left.\,C_{E_{1,2}}(c_1)C_{O_1}(c_2))\right|_{\,a=b=c=0} =0\,.
\\
\\
\ea \ee Substituting the functions (\ref{ferbos}), (\ref{express})
into (\ref{4.100}) one gets \be \label{4.1000} \ba{c} \dps
\frac{1}{2}\bar{k}^2\,{\rm
Res}_{\mu,\nu}\,((\Phi_1+\Phi_0)e^{\frac{1}{2}(\mu^{-1} \bar{u}_2
-\mu\bar{v}_2+\nu^{-1}\mu^{-1}\bar{k}
+\nu^{-1}\bar{v}_1+\nu\mu\bar{k}-\nu\bar{u}_1)})\frac{\d}{\d
c_2^\sigma} \times
\\
\\
\times {\rm
tr}(\epsilon_{O_2}{}^{\sigma\gamma(s)}_{\;\rho(s)}a_{3\;\gamma(s)}b_3^{\rho(s)}
\left.\,C_{E_{1,2}}(c_1)C_{O_1}(c_2))\right|_{\,a=b=c=0}
\\
\\
- \dps\frac{1}{2} \bar{k}^2\,{\rm Res}_{\mu,\nu}\, ((\Phi_1
+\Phi_0) \nu^{-1}\mu e^{\frac{1}{2}(\mu^{-1} \bar{u}_2
-\mu\bar{v}_2 +\nu\mu^{-1}\bar{k} +\nu\bar{v}_1
+\nu^{-1}\mu\bar{k} -\nu^{-1}\bar{u}_1)})\frac{\d}{\d
c_1^\sigma}\times
\\
\\
\times {\rm
tr}(\epsilon_{O_2}{}^{\sigma\gamma(s)}_{\;\rho(s)}a_{3\;\gamma(s)}b_3^{\rho(s)}
\left.\,C_{E_{1,2}}(c_1)C_{O_1}(c_2))\right|_{\,a=b=c=0} =0\,.
\\
\\
\ea \ee This is true provided that \be \label{phi0phi1}
\Phi_0=-\Phi_1\,. \ee Analogous analysis of the  terms with
$\epsilon_{O_1}$ in the higher spin transformation shows that the
invariance condition (\ref{SEtr}) is satisfied provided that
(\ref{1/4}) and (\ref{phi0phi1}) are true. The leftover ambiguity
in the coefficients (\ref{gab11})-(\ref{gab32}) reduces to an
overall factor $\Phi_0$ encoding the ambiguity in the
gravitational constant.

Thus, the action (\ref{acta}) is shown to properly describe the
higher spin ${\cal N}=1$ supersymmetric dynamics both at the free
field level and at the level of cubic interactions provided that
the coefficients of the bilinear form in $(\ref{acta})$ are fixed
according to (\ref{gab11})-(\ref{gab32}).

\section{Reduced Model}
\label{redmod}

So far we discussed  the $5d$ higher spin algebra $cu(1,1|8)$
being the centralizer of $N$ in the star product algebra. This
algebra is not simple as it  contains infinitely many ideals
$I_{P(N)}$ spanned by the elements of the form $P(N)\star F$ for
any $F \in cu(1,1|8)$ and any star-polynomial $P(N)$ \cite{FLA}.
In this section we focus on the algebra $hu_0 (1,1|8)$ that
results \cite{Vc} from  factoring out the maximal ideal
corresponding to $P(N) = N$. As we show, elements of this algebra
are spanned by the supertraceless multispinors. Thus  $hu_0
(1,1|8)$ describes the system of higher spin fields with every
supermultiplet emerging once. Note that the algebra $hu_0 (1,1|8)$
does not provide a maximal reduction of the original higher spin
algebra. The higher spin algebras with maximally reduced spectra
$ho_0(1,1|8)$ and its bosonic subalgebra $ho_0(1,0|8)$ were
discussed in \cite{SS1,Vc}.

We apply the approach elaborated for the pure bosonic system in
\cite{VD5} which consists of inserting a sort of projection
operator ${\cal M}$ to the quotient algebra into the action.
Namely, let ${\cal M}$ satisfy \be \label{condition} N\star {\cal
M} ={\cal M}\star N=0\,. \ee Having specified the "operator"
${\cal M}$ we write the action for the reduced system associated
with $hu_0 (1,1|8)$ by replacing the bilinear form in the action
with \be \label{bformc} {\cal A}(F,G) \to {\cal A}_0 (F,G)=  {\cal
A}(F,{\cal M}\star G)\,, \ee where ${\cal A}(F,G)$ corresponds to
the action describing the original (unreduced) higher spin
dynamics. To maintain gauge invariance we require ${\cal M}$ to
commute with elements of $cu(1,1|8)$ \be \label{final} F\star
{\cal M}= {\cal M}\star F\,,\qquad F \in cu(1,1|8)\,. \ee In fact,
this implies that ${\cal M}$ should be some star-function of $N$.
{}From the {\it C-invariance condition} it follows then \be {\cal
A}(F, {\cal M}\star G)={\cal A}(F\star {\cal M}, G)\,, \ee i.e.
the bilinear form in the action with ${\cal M}$ inserted remains
symmetric.

As a result, all terms proportional to $N$ do not contribute to
the action (\ref{bformc}) which therefore is defined on the
quotient subalgebra. The representatives of the quotient  algebra
$hu_0(1,1|8)$ are identified with the elements $F$ satisfying the
supertracelessness condition \be \label{straccon} P^-F=0\,. \ee
This allows one to require all fields in the expansion
(\ref{gf})-(\ref{gf4}) to be supertraceless. Indeed, by virtue of
(\ref{razl}) any polynomial $\tilde{F}(a,b,\psi,\bpsi|x)\in
cu(1,1|8)$ is equivalent to some $F$ satisfying (\ref{straccon})
modulo terms containing star products with $N$ which trivialize
when acting on ${\cal M}$. The star product $F\star G$ of any two
elements $F$ and $G$ satisfying the supertracelessness condition
does not necessarily satisfies the same condition, i.e. $P^-
(F\star G )\neq 0$ (otherwise the elements satisfying
(\ref{straccon}) would form a subalgebra rather than the  quotient
algebra). However the difference is again proportional to $N$ and
can be discarded inside the action built with the help of the
bilinear form ${\cal A}_0$.

To find explicit form of ${\cal M}$, one observes that any
star-function of $N$ is some (may be different) ordinary function
of $N$, i.e. \be \label{calm} {\cal M}(N) \equiv M(N)= M(a_\gamma
b^\gamma ) -M^\prime (a_\gamma b^\gamma )\psi\bpsi\,, \ee where
$M^\prime$ denotes a derivative of $M$. This is a simple
consequence of the fact that any such function has to commute with
the generators of $su(2,2|1)$. The later condition imposes some
first-order differential equations which are solved by an
arbitrary function of $N$.

The substitution of (\ref{calm}) into (\ref{condition}) results in
the second order differential equation \be \label{red72}
xM^{\prime\prime}(x)+3M^\prime(x)-4xM(x)=0\,,\qquad x\equiv
a_\gamma b^\gamma \,, \ee which admits a unique analytic solution
(up to a factor) \be \label{integral} M(x)=\int_0^1 d\tau\,{\rm
Res}_\nu\,e^{-\frac{1}{2}(\nu^{-1}+4\nu x^2 \tau)}\;. \ee
Equivalently \be \label{red71} M(x)=\sum_{n=0}^{\infty}\,\frac{x^{
2n}}{n!\,(n+1)!}\,. \ee

Having found the operator ${\cal M}$ we define the action for the
reduced system associated with $hu_0(1,1|8)$ in the form
(\ref{bformc}). Note that ${\cal A}(F,G)$ with inserted ${\cal M}$
according to (\ref{bformc}) is well-defined as a functional of
polynomial functions $F$ and $G$ because for polynomial $F$ and
$G$ only a finite number of terms in the expansion of ${\cal
M}(a_\alpha b^\alpha, \psi\bpsi)$ contributes.

The modification of the action according to (\ref{bformc}) does
not contradict to  the analysis of  section \ref{unred} where the
action (\ref{acta}) was claimed to be fixed unambiguously, because
in that analysis we have imposed the {\it factorization condition}
in the particular basis of higher spin fields thus not allowing
the transition to the invariant action (\ref{bformc}). The {\it
factorization condition} is relaxed in this section. All other
conditions, namely {\it C-invariance condition, extra field
decoupling condition} and the condition (\ref{SEtr}) remain valid.

\section{Conclusion}
\label{Conclusion}

In this paper we have analysed  cubic interactions in the theory
of higher spin fields in $AdS_5$ for the particular case of ${\cal
N}=1$ supersymmetry. It is shown that free field abelian  higher
spin gauge transformations admit such a deformation that the
constructed cubic action, that is general coordinate invariant and
contains gravity, remains invariant up to higher order terms.

Our conclusions are valid both for unreduced model based on
$cu(1,1|8)$ (every supermultiplet $(s,s-\frac{1}{2},s-1)$
determined by an integer highest spin $s=2,3,...,\infty \;$
appears in infinitely many copies) and for reduced model based on
$hu_0(1,1|8)$ symmetry in which every such supermultiplet appears
only once. In this respect our conclusions are different from
those of \cite{Fradkin:1991ps}, where it was argued that only
unreduced algebra $cu(1,1|8)$ admits consistent dynamics in the
framework of $4d$ higher spin conformal theory (although the two
models are different since the model of \cite{Fradkin:1991ps},
being a higher spin extension of the $4d$ $C^2$ gravity, contains
higher derivatives and ghosts, while our model in $AdS_5$ is
unitary in the physical space at least at the free field level).

Note that  the constructed higher-spin cubic vertices do not
exhaust all possible consistent supersymmetric higher-spin
interactions in $AdS_5$ in the order under consideration. One
reason for that is that we discard low-spin ($s\leq 1$)
interactions which  truncation is  consistent  in the cubic order
only. The study of the explicit form of cubic couplings of
particular higher spins in terms of physical fields is the
technically complicated problem requiring  full-scale
investigation which is beyond the scope of this paper. The
developed technics contains, however, all necessary ingredients
for the detailed analysis of the constructed interactions in terms
of physical fields which may be of interest in the context of
$AdS/CFT$ computations.

The generalization of the presented constructions to ${\cal N}\geq
2$ extended supersymmetry is not straightforward as it requires
mixed symmetry higher spin fields to be included \cite{Vc,VD5}.
The progress along this direction is hampered by lacking a
manifestly covariant Lagrangian description of massless gauge
fields of this type in $AdS_d$ with $d\geq 5$ even at the free
field level. The method employed in the present paper for
constructing higher spin cubic couplings is essentially based on
the Lagrangian formulation of  higher spin gauge field dynamics in
terms of appropriate connections \cite{V1,LV,vf}.  To proceed
towards ${\cal N}\geq 2$ an extension of this formalism to the
mixed symmetry fields at the Lagrangian level is  needed. Note
that the higher spin actions for mixed-symmetry higher spin fields
in anti-de Sitter space-time were built in different setups. In
\cite{Metsaev},  an explicit $AdS_5$ light-cone action describing
free mixed-symmetry fields has been constructed. In \cite{BS} an
approach to covariant description of an arbitrary representation
of $AdS_d$ algebra $o(d-1,2)$ is developed in the framework of the
radial reduction technique.

Also it would be useful to reformulate our results within a
superspace approach, which is shown to be a powerful tool in the
case of $4d$  free higher spin supermultiplets in Minkowski
spacetime \cite{Gates:nr}. Note that the off-shell superfield
realization  of ${\cal N}=1,2$ $AdS_4$ free higher spin massless
supermultiplets was given in \cite{Kuzenko:dm}. It would be
interesting to elaborate the superspace formulation for $d>4$ free
higher spin supermultiplets and extend the method of
\cite{Kuzenko:dm} to the study of the interaction problem.

\vskip0.5cm

{\bf Acknowledgments.} One of us (A.K.) would  like to thank
R.Metsaev, A.Segal and O.Sha\-ynk\-man for discussions. This
research was supported in part by INTAS Grant 99-1-590, RFBR Grant
02-02-17067 and RFBR Grant 01-02-30024. The work of A.K. is
partially supported by INTAS Grant No.00-00262 and the Landau
Scholarship Foundation, Forschungszentrum J\"ulich.

\section*{Appendix A. Compensator formalism in spinor notations}

For the reader's convenience in this appendix we collect some
useful formulae on the compensator formalism in spinor notations
developed in \cite{VD5}.

A $o(6)$  complex vector $V^A$ ($A=0,...,5$) is equivalent to the
antisymmetric $sl_4$ bispinor $V^{\alpha\beta}=-V^{\beta\alpha}$
having six independent components (equivalently, one can use
$V_{\alpha\beta} = \half
\varepsilon_{\alpha\beta\gamma\delta}V^{\gamma\delta}$ where
$\varepsilon_{\alpha\beta\gamma\delta}$ is the $sl_4$ invariant
totally antisymmetric tensor ($\varepsilon_{1234} =1$)). A
$o(4,2)$ real vector $V^A$ is described by the antisymmetric
bispinor $V^{\alpha\beta}$ satisfying the reality condition \be
\overline{V}^{ \gamma \delta} C_{\gamma \alpha} C_{\delta \beta}=
\half \varepsilon_{\alpha\beta\gamma\delta}V^{\gamma\delta}\,. \ee
One can see that the invariant norm of the vector \be \label{norm}
V^2 = {V}_{\alpha\beta} V^{\alpha\beta} \ee has the signature
$(++----)$. The vectors with $V^2 >0$ are time-like while those
with $V^2 <0$ are space-like. To perform a reduction of the
representations of the $AdS_5$ algebra $su(2,2)\sim o(4,2)$ into
representations of its Lorentz subalgebra $o(4,1)$ we introduce a
$su(2,2)$ antisymmetric compensator $V^{\alpha\beta}$ with
$V^2>0$. The Lorentz algebra is identified with its stability
subalgebra. (Let us note that $V^{\alpha\beta}$ must be different
from the form $C^{\alpha\beta}$ used in the definition of the
reality conditions (\ref{inv}) - (\ref{reco}) since the latter is
space-like and therefore has $sp(4; R )\sim o(3,2)$ as its
stability algebra.)

Using that the total antisymmetrization over any four indices is
proportional to the $\varepsilon$ symbol, we normalize
$V^{\alpha\beta}$ so that \be \label{vno1} V_{\alpha\beta}
V^{\alpha\gamma} = \delta_\beta{}^\gamma \,,\qquad
V_{\alpha\beta}=\half
\varepsilon_{\alpha\beta\gamma\delta}V^{\gamma\delta}\,, \ee \be
\varepsilon_{\alpha\beta\gamma\delta} =
V_{\alpha\beta}V_{\gamma\delta} +V_{\beta\gamma}V_{\alpha\delta}
+V_{\gamma\alpha}V_{\beta\delta}\,, \ee \be
\varepsilon^{\alpha\beta\gamma\delta} =
V^{\alpha\beta}V^{\gamma\delta} +V^{\beta\gamma}V^{\alpha\delta}
+V^{\gamma\alpha}V^{\beta\delta}\,. \ee The gravitational fields
are identified with the gauge fields taking values in the $AdS_5$
algebra $su(2,2)$ \be \Omega =\Omega{}^\alpha{}_\beta a_\alpha
b^\beta\,. \ee The invariant definitions of the frame field and
Lorentz connection for a $x-$dependent compensator
$V^{\alpha\beta} (x)$ are \be \label{h} E^{\alpha\beta} =D
V^{\alpha\beta} \equiv dV^{\alpha\beta} + \Omega{}^\alpha{}_\gamma
V^{\gamma\beta} + \Omega{}^\beta{}_\gamma V^{\alpha\gamma}
\,,\qquad \ee \be \Omega^{L\,\alpha}{}_\beta =
\Omega{}^\alpha{}_\beta +\half E^{\alpha\gamma} V_{\gamma
\beta}\,. \ee The normalization condition (\ref{vno1}) implies \be
\label{tr11} E_{\alpha\beta} =-D V_{\alpha\beta}\,, \qquad
E_\alpha{}^\alpha =0\,. \ee $AdS_5$ background geometry is defined
by zero-curvature condition \be R{}^\alpha{}_\beta \equiv d
\Omega^\alpha{}_\beta +\Omega^\alpha{}_\gamma\wedge
\Omega^\gamma{}_\beta =0\,, \ee which decomposes into Lorentz
components as \be \label{srt} R^{L}{}^\alpha{}_\beta \equiv
d\Omega^{L}{}^\alpha{}_\beta +\Omega^{L}{}^\alpha{}_\gamma\wedge
\Omega^{L}{}^\gamma{}_\beta +\frac{1}{4}E^\alpha{}_\gamma\wedge
E^\gamma{}_\beta=0\,, \ee \be \label{Tor} T{}^{\alpha\beta} \equiv
d E^{\alpha\beta} + \Omega{}^\alpha{}_\gamma \wedge
E^{\gamma\beta} + \Omega{}^\beta{}_\gamma \wedge
E^{\alpha\gamma}=0 \,. \ee (\ref{Tor}) is the  conventional
zero-tension condition, while the equation  (\ref{srt})  requires
the geometry to be anti-de Sitter.

The nondegeneracy condition implies
 that $ E^{\alpha \beta}$ spans a basis of the
$5d$ 1-forms. The basis $p$-forms $E_p$ can be realized as \be
\label{3may} E_2^{\alpha\beta}=E_2^{\beta\alpha}
=E^{\alpha}{}_\gamma \wedge E^{\beta\gamma}\,, \ee \be
E_3^{\alpha\beta}=E_3^{\beta\alpha} =E_2^{\alpha}{}_\gamma \wedge
E^{\beta \gamma }\,, \ee \be E_4^{\alpha\beta}=-E_4^{\beta\alpha}
=E_3^{\alpha}{}_\gamma \wedge E^{\beta \gamma}\,, \ee \be
\label{4may} \label{E5} E_5=E_4^{\alpha}{}_\gamma \wedge
E_{\alpha}{}^{\gamma}\,. \ee The following useful relationships
hold as a consequence of the facts that $5d$ spinors have four
components and the frame field is traceless \be \label{id1}
E^{\alpha\beta}\wedge E^{\gamma\delta} = \half (
V^{\alpha\gamma}E_2^{\beta\delta}
-V^{\beta\gamma}E_2^{\alpha\delta}
-V^{\alpha\delta}E_2^{\beta\gamma}
+V^{\beta\delta}E_2^{\alpha\gamma} )\,, \ee \be
E_2^{\alpha\beta}\wedge E^{\gamma\delta} = -\frac{1}{3} (
V^{\alpha\gamma}E_3^{\beta\delta}+
V^{\beta\gamma}E_3^{\alpha\delta}
-V^{\beta\delta}E_3^{\alpha\gamma}
-V^{\alpha\delta}E_3^{\beta\gamma}
+V^{\gamma\delta}E_3^{\alpha\beta} )\,, \ee \be
E_{4\alpha}{}^\alpha =0\,, \ee \be E^{\alpha\beta}\wedge E_3
^{\gamma\delta} = -\frac{1}{4} (V^{\alpha\gamma}E_4^{\beta\delta}
-V^{\beta\gamma}E_4^{\alpha\delta}
+V^{\alpha\delta}E_4^{\beta\gamma}
-V^{\beta\delta}E_4^{\alpha\gamma} )\,, \ee \be \label{id2}
E_4^{\alpha\beta} \wedge E^{\gamma\delta} =
-\frac{1}{20}(2V^{\alpha\gamma} V^{\beta\delta} -
2V^{\alpha\delta}V^{\beta\gamma} -V^{\alpha\beta}V^{\gamma\delta}
)E_5 \,. \ee The background frame field and Lorentz connection are
denoted $h=h^\alpha{}_\beta a_\alpha b^\beta$ and
$\Omega_0^L=\Omega_0^{L\,\alpha}{}_\beta a_\alpha b^\beta$,
respectively. Vacuum values of the $p$-forms $E_p^{\alpha\beta}$
(\ref{3may})-(\ref{4may}) are denoted $H_p^{\alpha\beta}$.

\section*{Appendix B. Trace and supertrace decompositions in $cu(1,1|8)$. }

In this appendix we show how to derive the expansions
(\ref{tr1})-(\ref{tr2}). We start with \be \label{STRDB}
\Omega(a,b,\psi,\bpsi|x) =
\sum_{k=0}^{\infty}\,\sum_{s=1}^{\infty}\,\chi(k,\,s)\,(P^+)^k
\;\Omega^{k,\,s+1}(a,b,\psi,\bpsi|x)\,, \ee where $\chi(k,\,s)$
are some coefficients, $s+1$ denotes highest integer spin in a
supermultiplet, $\Omega^{k,s+1}$ satisfy $P^0\Omega^{k,\,s+1}
=(2s+3)/4\,\Omega^{k,\,s+1}$ and are supertraceless \be
\label{STRB} P^-\,\Omega^{k,\,s+1}(a,b,\psi,\bpsi|x) =0\,. \ee The
solution of (\ref{STRDB})-(\ref{STRB}) is given by (\ref{AB1}),
(\ref{AB2}). Taking into account that \be \label{AB3}
(P^+)^k=(T^+)^k-k(T^+)^{k-1}\psi\bpsi \ee (the usual product here
should not be confused with the star product) after some algebra
one gets from (\ref{STRDB}) \be \label{su_base} \ba{c} \dps
\Omega(a,b,\psi,\bpsi|x) =
\sum_{k,s=0}^{\infty}\beta(k,s)(T^+)^k\Omega^{k,s+1}_{E_1}(a,b|x)+
\sum_{k,s=0}^{\infty}\rho(k,s)(T^+)^k
\Omega^{k,s+1}_{E_2}(a,b|x)\psi\bpsi
\\
\\
+\dps\sum_{k,s=0}^{\infty}\chi(k,s+1)(T^+)^k
\Omega^{k,s+3/2}_{O_1}(a,b|x)\psi
+\sum_{k,s=0}^{\infty}\chi(k,s+1)(T^+)^k
\Omega^{k,s+3/2}_{O_2}(a,b|x)\bpsi

\;, \ea \ee where \be \label{transit} \beta(k,s)
\Omega^{k,s+1}_{E_1}(a,b|x)=
\theta(k-1)\chi(k-1,s+1)\tilde{\Omega}^{k-1,s+1}_{E_2}(a,b|x) +
\theta(s-1)\chi(k,s)\tilde{\Omega}^{k,s+1}_{E_1}(a,b|x)\;, \ee \be
\label{transit2} \ba{c} \dps \rho(k,s)
\Omega^{k,s+1}_{E_2}(a,b|x)=
\chi(k,s+1)\tilde{\Omega}^{k,s+1}_{E_2}(a,b|x)
\\
\\
-\dps\theta(s-1)(k+1)\chi(k+1,s)\tilde{\Omega}^{k+1,s+1}_{E_1}(a,b|x)
+ \frac{k}{2s+5} \chi(k,s+1)\tilde{\Omega}^{k,s+1}_{E_2}(a,b|x)\;,

\ea \ee \be \Omega^{k,s+3/2}_{O_1}(a,b|x) =
\tilde{\Omega}^{k,s+3/2}_{O_1}(a,b|x)\,,\quad
\Omega^{k,s+3/2}_{O_2}(a,b|x) =
\tilde{\Omega}^{k,s+3/2}_{O_2}(a,b|x)\,. \ee All multispinors on
the l.h.s. of (\ref{su_base}) are traceless as a consequence of
(\ref{AB2}). We see that supertraceless and traceless bases are
related by a finite linear field redefinition.

\end{document}